\documentclass[a4paper,11pt]{article}
\pdfoutput=1 

\usepackage{jcappub} 
\usepackage{empheq}

\usepackage[normalem]{ulem}
\usepackage{hyperref}
\usepackage{multirow}
\usepackage{array}

\usepackage{graphicx}
\usepackage{graphbox}

\usepackage{dcolumn}
\usepackage{bm}

 \usepackage{amssymb,amsmath}

 \usepackage{stackengine,graphicx}
 \usepackage[normalem]{ulem}

\title{Thermodynamics of \texorpdfstring{$f(R)$}{fR} Gravity: The Double Well Potential Case}

\author[a]{C.D. Peralta}
\author[b]{S.E. Jor\'as}

\affiliation[a]{
Centro de Investigaciones en Ciencias B\'asicas y Aplicadas, Universidad Antonio Nari\~no,
Cra 3 Este \# 47A-15, Bogot\'a D.C. 110231, Colombia
}
 \affiliation[b]{
 Instituto de F\'\i sica, Universidade Federal do Rio de Janeiro,\\
 CEP 21941-972 Rio de Janeiro, RJ, Brazil
}

 \emailAdd{cperalta294@uan.edu.co}
 \emailAdd{joras@if.ufrj.br}

\abstract{
In this work we further extend the analysis of $f(R)$ theories of gravity in the metric formalism under the approach of a Thermodynamics analogy, proposed in \cite{Peralta}. Here we assume a double-well inflationary potential in the Einstein frame and obtain a parametric form of $f(R)$ in the corresponding Jordan frame. 

The whole Thermodynamics picture then follows: an equation of state, binodal and spinodal curves, phase transition, critical quantities (pressure, volume and temperature), entropy jumps, specific-heat divergence (and the corresponding critical exponent) and a butterfly catastrophe.}

\begin{document}

\maketitle
\flushbottom

\section{Introduction}

In this paper we focus on $f(R)$ theories \cite{DeFelice:2010aj,Capozziello:2011et,Nojiri:2017ncd}  --- nonlinear functions of the Ricci scalar $R$ defined, as usual, in the Jordan Frame (JF). We follow the metric formalism, which features an extra degree of freedom (d.o.f), as we will briefly review.
It is well known that, upon a suitable conformal transformation (as we will also recall below), the modified gravitational Lagrangian assumes the usual Einstein-Hilbert form and the extra d.o.f. is materialized as a scalar field --- for obvious reasons, this is the so-called Einstein Frame (EF). \textcolor{black}{See, for instance, Ref.~\cite{2006PhLB..634...93C} for a discussion on how to determine the ``true physical frame''.}

Here, we will follow the same path, but in the opposite direction: we start from a double well potential (DW) $V_E(\phi)=(\phi^2 -a^2)^2+\Lambda$  with an {\it ad-hoc} Cosmological Constant $\Lambda$ in the EF (with also standard slow-roll initial conditions) and investigate the corresponding $f(R)$ in the JF. The introduction of  $\Lambda$ will  lead us to a full thermodynamical approach to $f(R)$ theories, shedding some light on the evolution of the system in both frames --- interesting results are still obtained even for the plain $\Lambda=0$ case.

We will now briefly review the aforementioned conformal transformation and the mapping from the quantities defined in one frame to their corresponding {\it Doppelg\"angers} in the other frame. 

\section{Conformal Transformation and the Inverse Problem}
\label{conformal}

From now on, the super(sub)scripts ``$^E$",  ``$^J$" indicate the frame (Einstein and Jordan, respectively) where the quantity is defined. We drop the subscript in $R_J\equiv R$ (and in $\phi_E\equiv \phi$ --- see below) to avoid excessive cluttering of the equations. 

We write the modified gravitation Lagrangian in JF (in the vacuum, i.e, no matter/radiation fields) as
\begin{equation}\label{Lvacuum}
L_J = \sqrt{-g^J} f(R),
\end{equation}
where $g^J\equiv \det (g^J_{\mu\nu})$ is the determinant of the metric in the JF. General Relativity (GR) with a cosmological constant $\Lambda$ would correspond to a linear $f(R) = R - 2\Lambda$, and vice-versa. The standard variational procedure in the metric formalism yields fourth-order equations for the metric \cite{Sotiriou:2008rp}
\begin{equation}\label{EqfieldfR}
 R_{\mu\nu} f' -\frac12  g^J_{\mu\nu} f + g^J_{\mu\nu}\, \Box f' - \nabla_{\mu}\nabla_{\nu} f' = 0,
\end{equation}
where $f' \equiv {\rm d} f/{\rm d} R$. 

One then introduces the new pair of variables $\{g^E_{\mu\nu},p\}$, related to $g^J_{\mu\nu}$ (and to its derivatives) by a conformal transformation from the JF to the EF \textcolor{black}{\cite{Maeda, Magnano:1993bd,Wald:1984rg}}:
\begin{equation}
g^{E}_{\mu\nu} \equiv \Omega^2(x^\alpha) \, g^{J}_{\mu\nu}\, , \quad {\rm where} \quad  \Omega^2 \equiv p  \equiv f'(R).
\end{equation}
We now define $R(p)$ as a solution of the equation $f'[R(p)] - p = 0$. This procedure corresponds to a standard Legendre Transformation. As such, the expression $R(p)$ is uniquely defined as long as $f''\equiv d^2f/d R^2$ has a definite sign. Nevertheless, it is possible to write a unique expression for $R(\phi)$ --- see Eq.~(\ref{Rphi}) below --- which holds across the branches where $f''(R)$ has different signs, and yields smooth functions $R(t)$ and $\phi(t)$ across the branches. 
A scalar field  $\phi_E\equiv \phi$  (dropping the subscript) is traditionally defined in the EF by $p \equiv \exp{(\beta \,\phi)}$, with $\beta \equiv \sqrt{2/3}$. The Lagrangian (\ref{Lvacuum}) can then be recast in a more familiar form:
\begin{equation} \label{LagrangianE2}
L_E = \sqrt{-g^E} \Bigg[R_E - g_E^{\mu\nu} \phi_{,\mu} \phi_{,\nu} - 2V_E(\phi)\Bigg],
\end{equation}
where $R_E$ is the Ricci scalar obtained from $g^E_{\mu\nu}$. In other words, in the EF, the gravitational dynamics is set by a GR-like term ($R_E$) and the field $\phi$ is an ordinary minimally-coupled massive scalar field subject to the potential \cite{Magnano:1993bd}
\begin{equation}
V_E(\phi) \equiv \frac{1}{2p^2}\Big\{p R[p(\phi)] - f[R(p(\phi))] \Big\}
\end{equation}
which is completely determined by the particular $f(R)$ chosen. 

In the present work we start by examining the inverse problem: from a scalar field $\phi$ and its potential $V_E(\phi)$,  we map $L_E$ in Eq.~(\ref{LagrangianE2}) onto the corresponding $L_J$ in Eq.~(\ref{Lvacuum}). 
Following a previoulsy established procedure \cite{Magnano:1993bd}, one arrives at the following parametric expressions:
\begin{align}
\label{fphi}
f(\phi) &= {\rm e}^{2 \beta \phi} \left[2V_E(\phi) + 2 \beta^{-1}  \frac{{\rm d} V_E(\phi)}{d\phi} \right] \quad {\rm and}\\
\label{Rphi}
R(\phi) &= {\rm e}^{\beta \phi} \left[4 V_E(\phi) + 2 \beta^{-1} \frac{{\rm d} V_E(\phi)}{d\phi} \right].
\end{align}
We will apply the above equations to the DW potential for a scalar field, to which we also add an {\it ad hoc} Cosmological Constant $\Lambda$: 
\begin{equation}\label{VE}
V_E(\phi) \equiv  \frac{m^2_\phi}{8a^2}\,  (\phi^2 -a^2)^2 + \Lambda,
\end{equation}
where $a$ is the vacuum expectation value, which rescales the effective cosmological constant in the JF (see discussion below).
One might argue that the insertion of $\Lambda$ goes completely against the reasoning of modifying GR but, for now, $\Lambda$ is written just for the sake of completeness. As we will see later on, it will turn out to be a key ingredient for the thermodynamic interpretation. 
Besides, here we focus on the primordial universe, where the accelerated expansion is {\it not} generated by a $\Lambda$-like term. Still, even the standard case  $\Lambda=0$ yields very interesting results, as we will see later on. 

\begin{figure}
\center
  \includegraphics[width=0.35\textwidth]{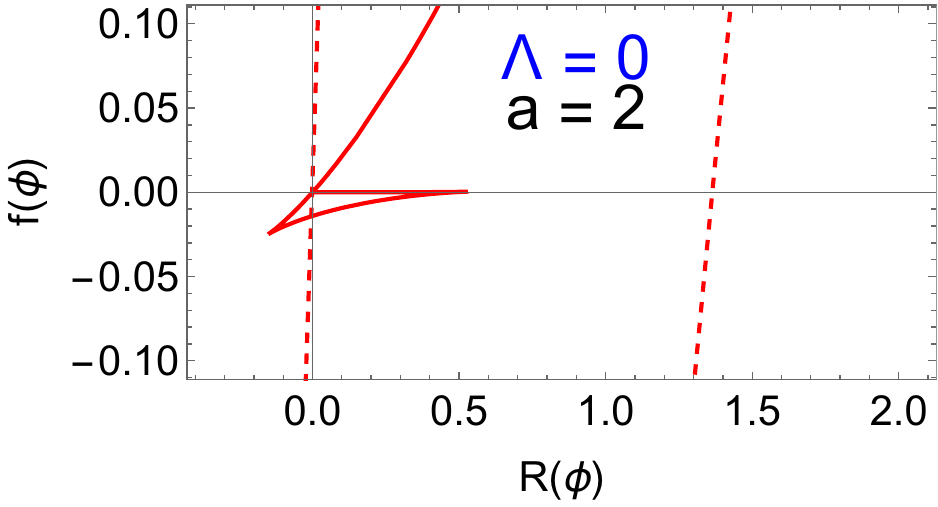}
  \includegraphics[width=0.35\textwidth]{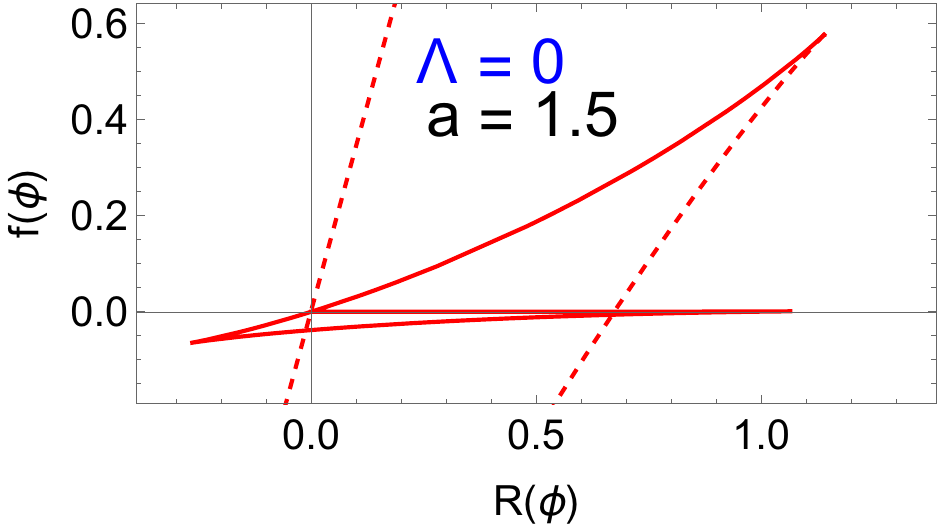}
   \includegraphics[width=0.35\textwidth]{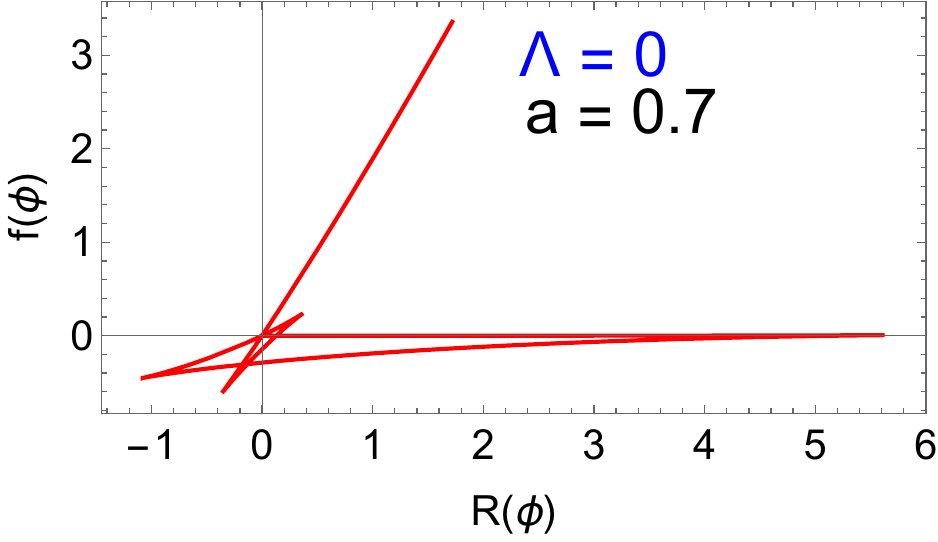}
    \includegraphics[width=0.35\textwidth]{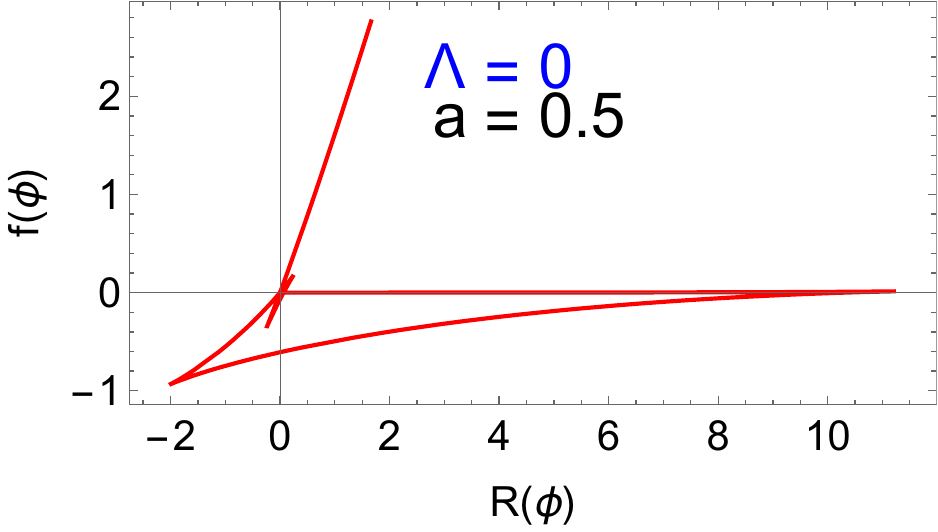}
\caption{Parametric plots of $f(R)$ given by Eqs. (\ref{fVE}, \ref{RVE}). Dynamics of parameter $a$ fixing $\Lambda =0$.  \textbf{In the top panels}, the potential barrier is too high and the evolution of the system reproduces the result for a single well \cite{Peralta}. The red-dashed lines show the path that {\it would} be followed if the field could go over such barrier. \textbf{In the lower panels}, the potential barrier is low enough for the field to reach the second well and then it presents this new second behaviour on $f(R)$.}
\label{Butterfly1}
\end{figure}

\begin{figure}
\center
  \includegraphics[width=0.35\textwidth]{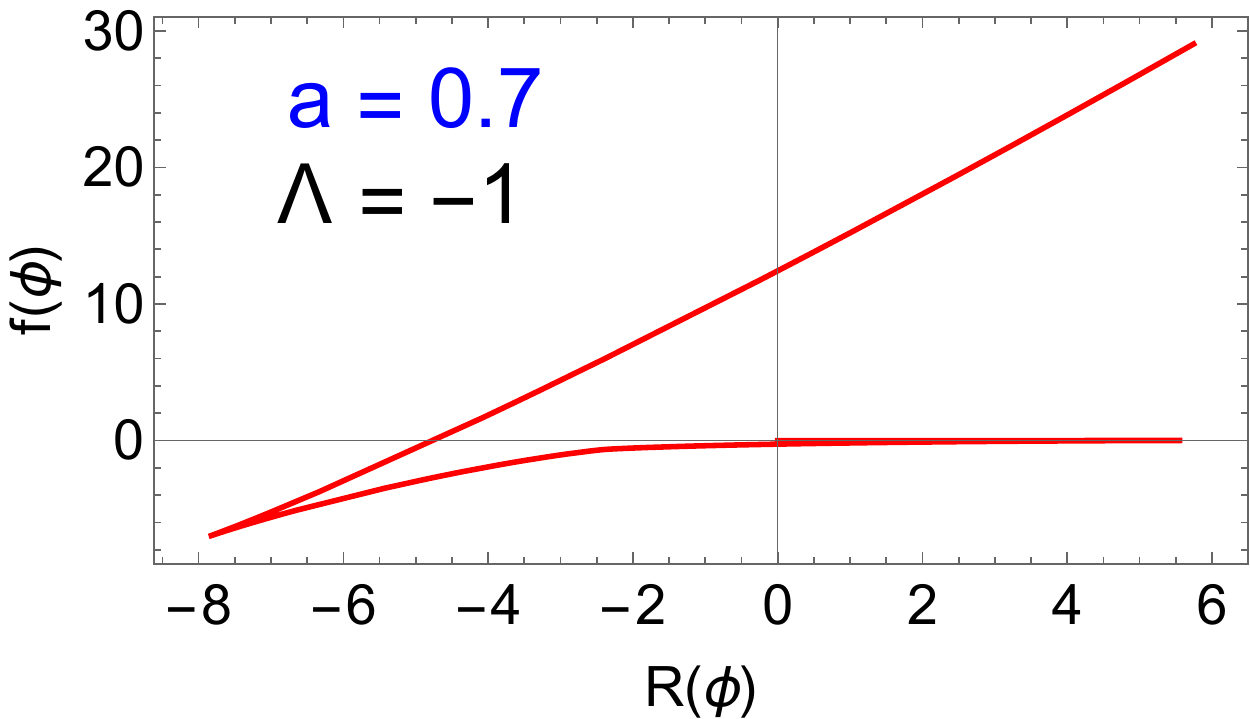}
  \includegraphics[width=0.37\textwidth]{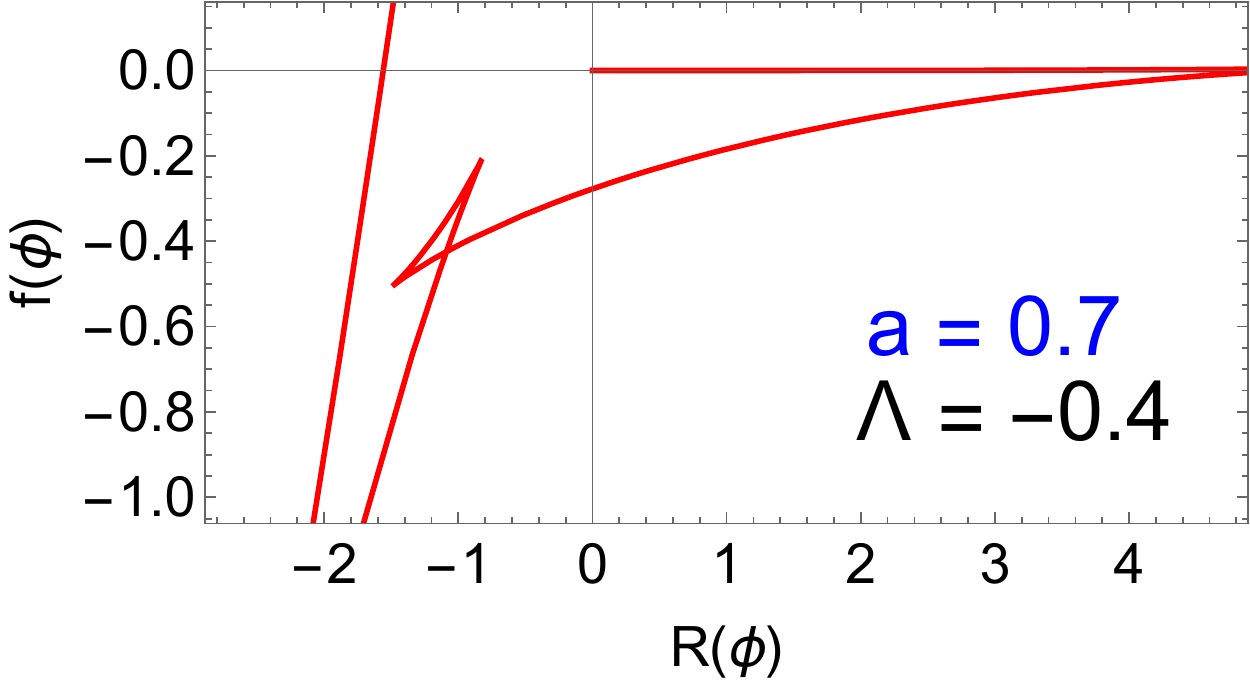}
   \includegraphics[width=0.37\textwidth]{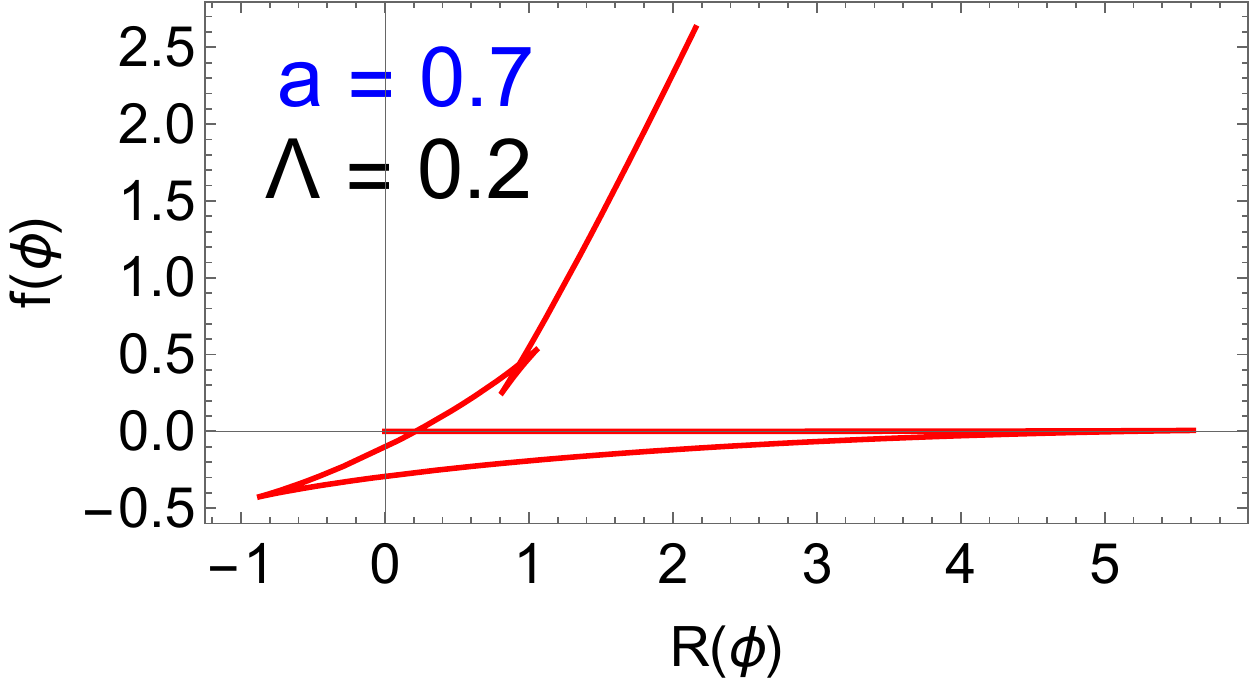}
    \includegraphics[width=0.35\textwidth]{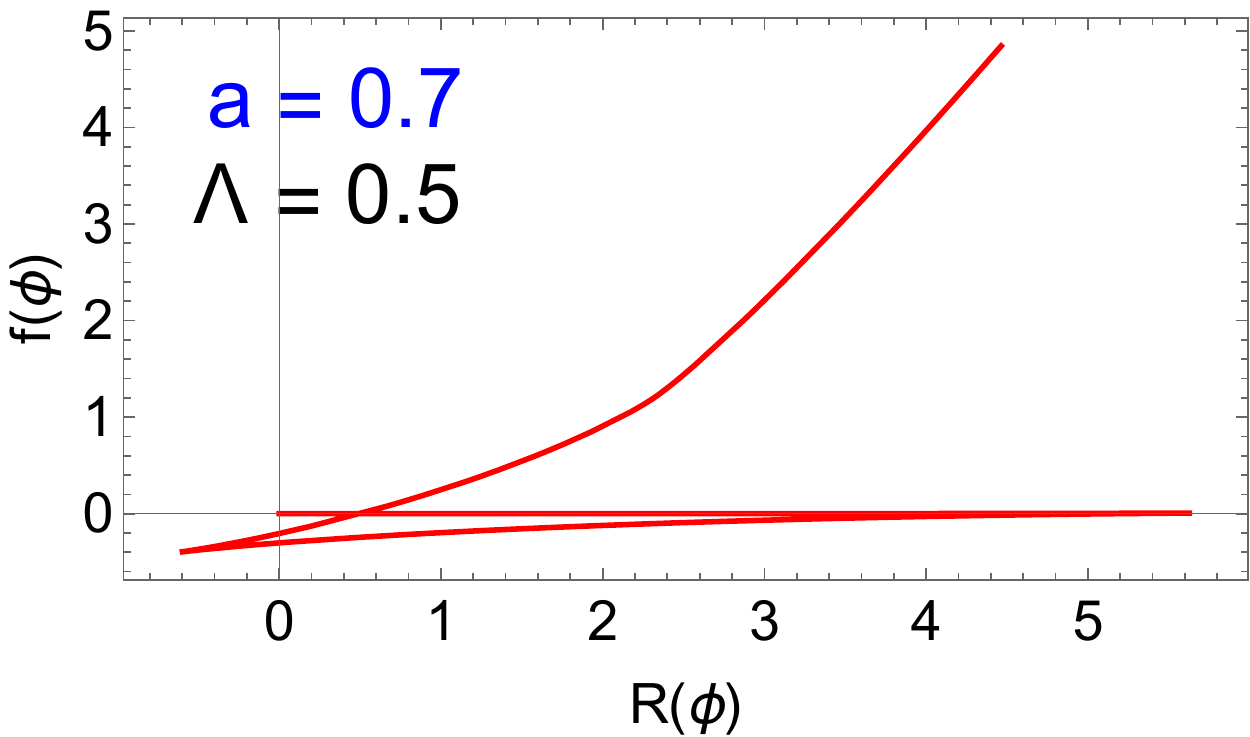}
    \includegraphics[width=0.37\textwidth]{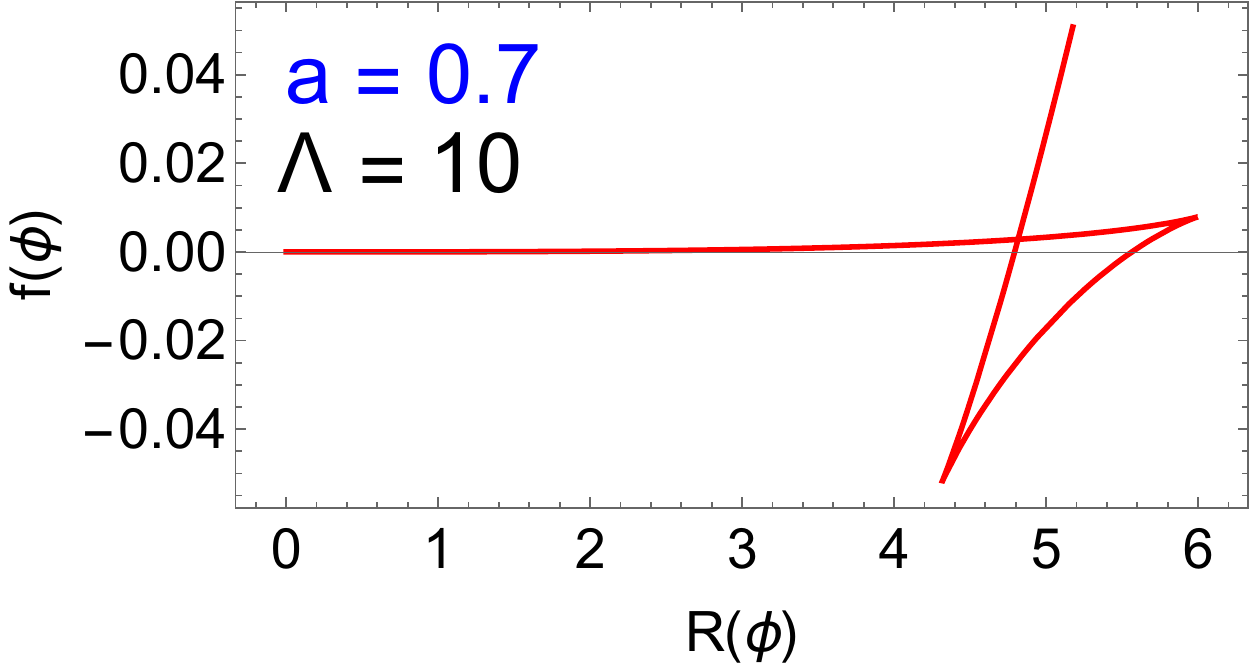}
    \includegraphics[width=0.36\textwidth]{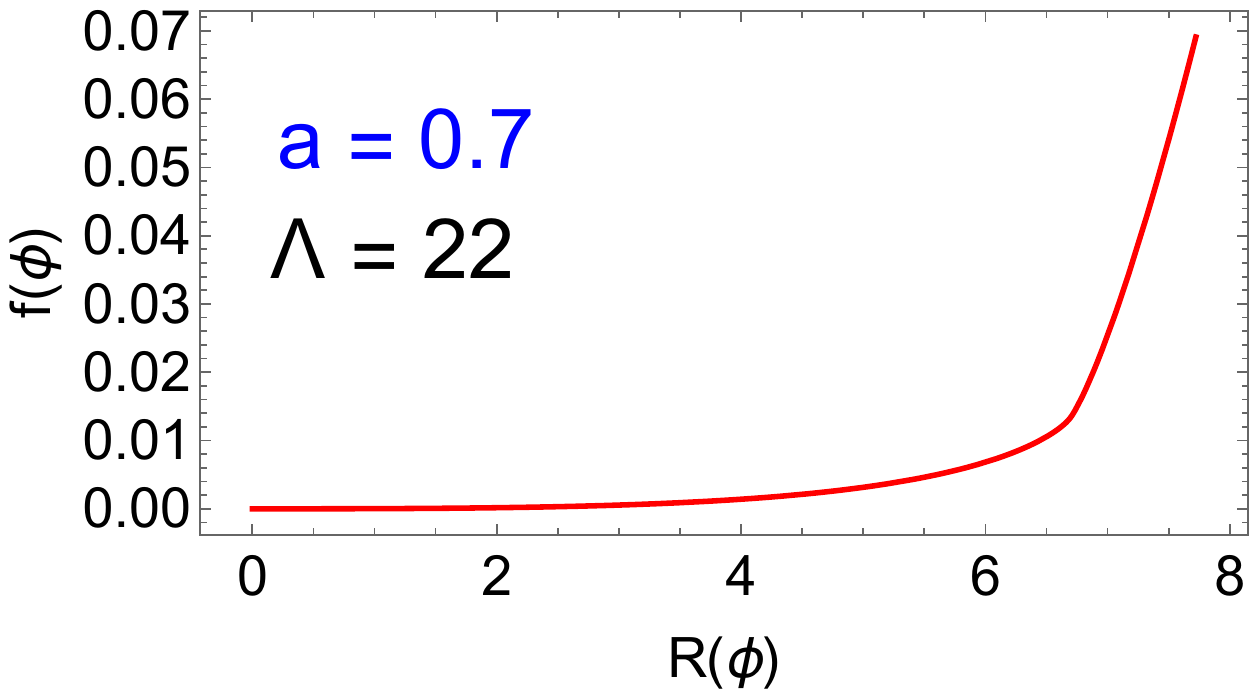}
\caption{Parametric plots of $f(R)$ given by Eqs. (\ref{fVE}, \ref{RVE}) for increasing values of the parameter $\Lambda$ with  fixed $a=0.7$ (for which the field ends at the bottom of the second well). In all plots, there is a branch close to the horizontal axis that can be clearly seen only in the final panel. There are three critical values of $\Lambda$: \textbf{Top panels:} At $\Lambda \approx  -1.05$, a new unstable branch appears with positive concavity, yielding a five-branch structure. \textbf{In center panels:} At $ \Lambda \approx 0.4$ the five-branch structure ends and the system has three branches again. \textbf{In low panels:} At $\Lambda \approx 21.04$ the three-branch structure turns into to a single-branch one.}
\label{Butterfly2}
\end{figure}

Eqs. (\ref{fphi}) and (\ref{Rphi}) then yield the corresponding parametric form of $f(R)$:
\begin{align}
\label{fVE}
f(\phi) &= e^{2 \beta  \phi } \left[2 \left(\frac{m^2_\phi \left(\phi ^2-a^2\right)^2}{8 a^2}+\Lambda \right) + \frac{m^2_\phi \phi  \left(\phi ^2-a^2\right)}{a^2 \beta }\right]\\
\label{RVE}
R(\phi) &= e^{\beta  \phi } \left[4 \left(\frac{m^2_\phi \left(\phi ^2-a^2\right)^2}{8 a^2}+\Lambda \right)+\frac{m^2_\phi \phi  \left(\phi ^2-a^2\right)}{a^2 \beta }\right],
\end{align}
which we plot in Figs.~\ref{Butterfly1} and \ref{Butterfly2} for different values of free parameters $a$ and $\Lambda$, In all panels, $f'>0 \, \forall R$. The field $\phi$ and $a$ are given in Planck-Mass ($M_{\rm pl}$) units, $R$ and $\Lambda$ are given in $M_{\rm pl}^4$. We used $m_\phi=1 M_{\rm pl}$. 

Since the height of the central potential barrier $V_E(\phi=0)=m^2_\phi a^2/8$ (obviously) depends on $a$, there is a critical value $a_c \approx 0.81$ below which the initial mechanical energy of the scalar field $\phi$ (determined by requiring slow-roll initial conditions) is high enough to allow it to go above the central barrier and the field will end up oscillating in the second well (lower panels in Fig.~\ref{Butterfly1}).

%
%


We remind the reader that it is not necessary to have a large $R$ when the density is large (e.g, at early times), as in GR, because there is no algebraic relation between $R$ and $T$. Instead, here we have a differential equation, where $\rho$ is just the source for the evolution of $R$. Actually, we have no $\rho$ (so, no source term for $R$) and, indeed, $R\approx 0$ at early times but it increases during inflation and eventually it oscillates around $R\approx 0$. In the absence of matter/radiation, $R$ would identically vanish in GR.

The usual constraint on the second derivative of $f(R)$ --- $ d^2f/dR^2>0$ --- is necessary so that  empty background solutions are stable \cite{DeFelice:2010aj}. In the present work (as well as in Ref.~\cite{Peralta}), the system does feature such instabilities, but they are only temporary.
%
%
\section{Numerical Analysis for the DW potential}
\subsection{Einstein Frame (EF)}
From now on, we will investigate the potential given in Eq.~(\ref{VE}) as a inflationary potential in the EF --- initially, we will keep $a=0.7$ and $\Lambda=0$, except when necessary for a cleaner picture and noted so. 

First of all, we have to determine the time evolution of $R(t)$ and $\phi(t)$. We recall that throughout this paper there is no matter nor radiation, since the $\phi$ field is actually a gravitational d.o.f., expressed as a scalar field in the EF. In GR, that would imply $R=0 \, \forall \, t$. In $f(R)$ theories, on the other hand, $R$ has a dynamical behavior of its own. Here, it suffices to use $R[\phi(t)]$ (defined in the JF) from Eq.~(\ref{RVE}) and  $\phi(t)$ (in the EF) from the standard equation of motion for a scalar field in an expanding homogeneous spacetime:
\begin{equation}
 \ddot{\phi}(t) + 3 H(t) \dot{\phi}(t) +  V_E'[\phi(t)]=0,
 \label{eqphi}
\end{equation}
where $V_E'\equiv dV_E/d\phi$ and $H^2(t) = \{ \dot{\phi}(t)^2/2 + V_E[\phi(t)]\}/3$. The initial conditions for the numerical solution of Eq.~(\ref{eqphi}) are the standard ones in the slow-roll approximation \cite{Linde:2007fr}:
$\phi(0) \approx - 100$ and  $\dot \phi(0) \approx 116.61$,  which correspond to $ R(0)  \approx 3.45 \times 10^{-28}$ and $\dot R(0) \approx 2.67 \times 10^{-28}$. \footnote{Where $\phi$ is given in Planck-Mass ($M_{\rm pl}$) units, $R$ is given in $M_{\rm pl}^4$,
and $N=60$ is the number of efolds.} 
We point out that the slow roll is an attractor in the double-well inflation \cite{RODRIGUES2021136156} so that the initial conditions do not need to be fine tuned.

\begin{figure}
\center
\includegraphics[width=0.3\textwidth]{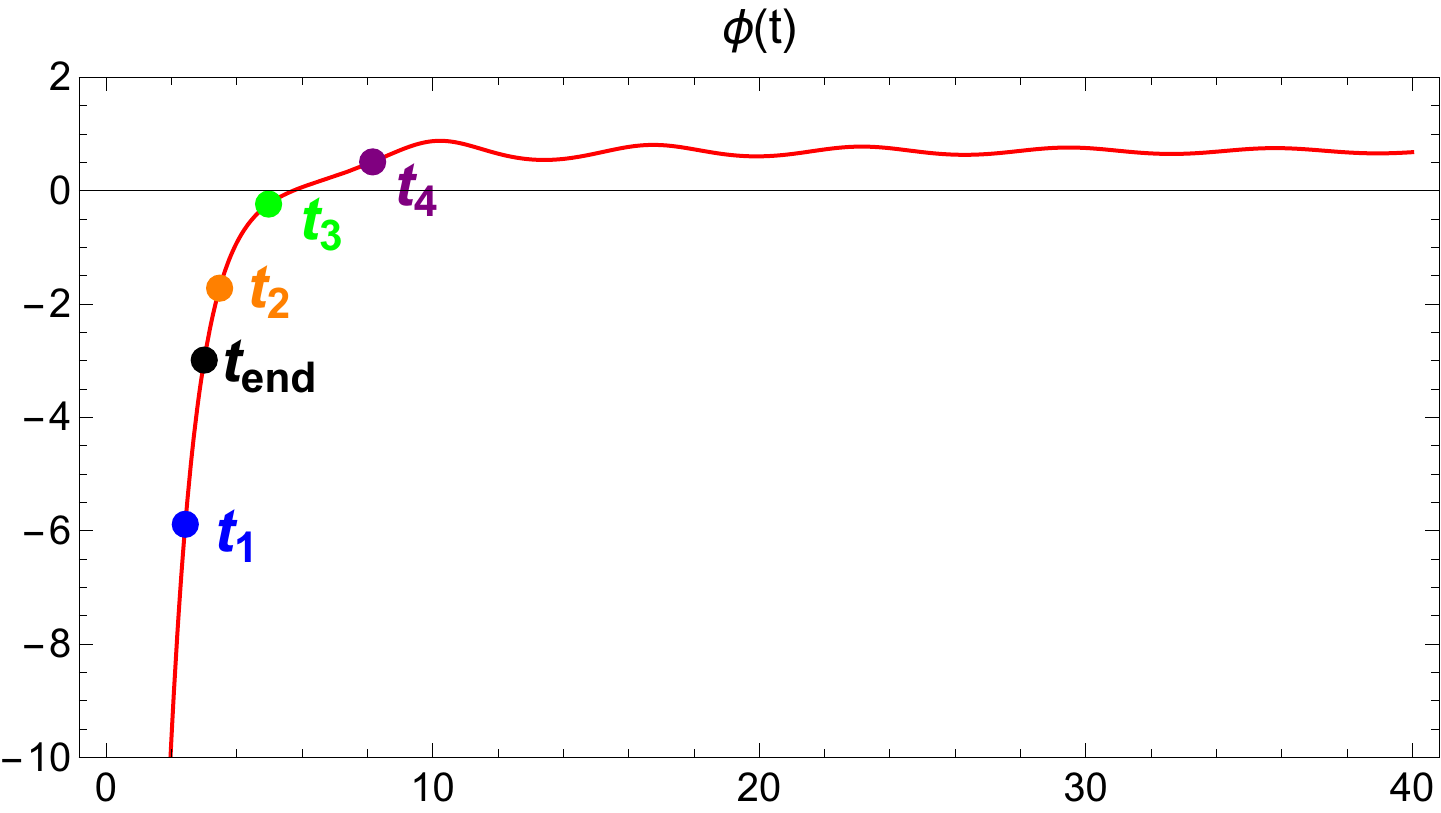}
\includegraphics[width=0.3\textwidth]{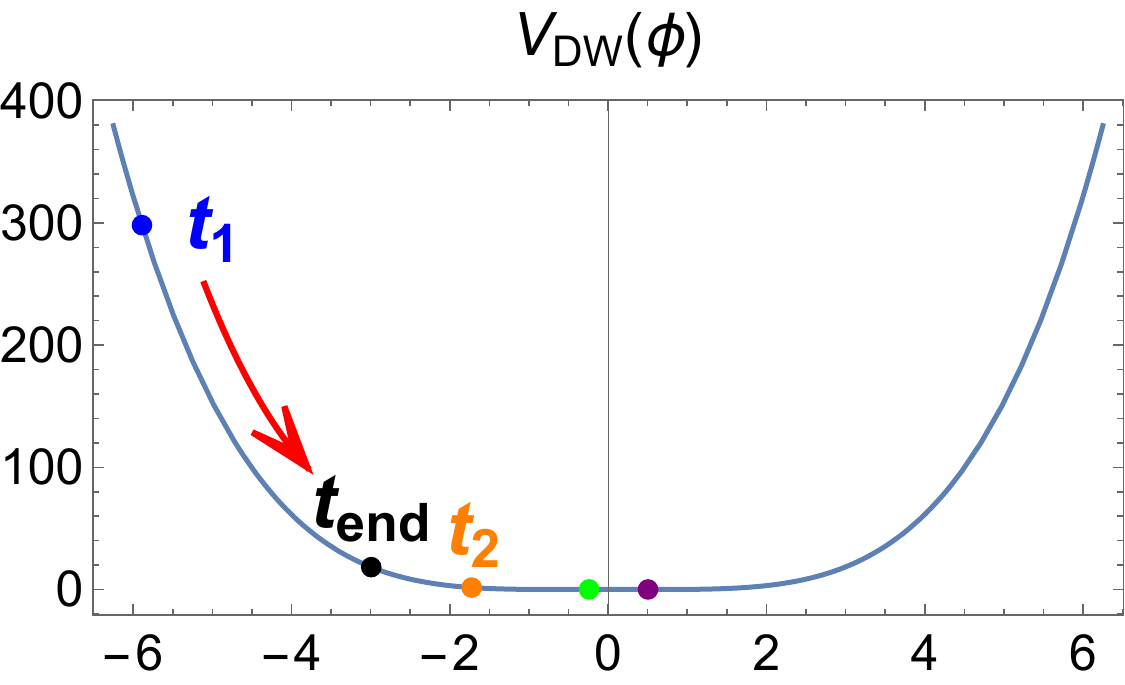}
\includegraphics[width=0.3\textwidth]{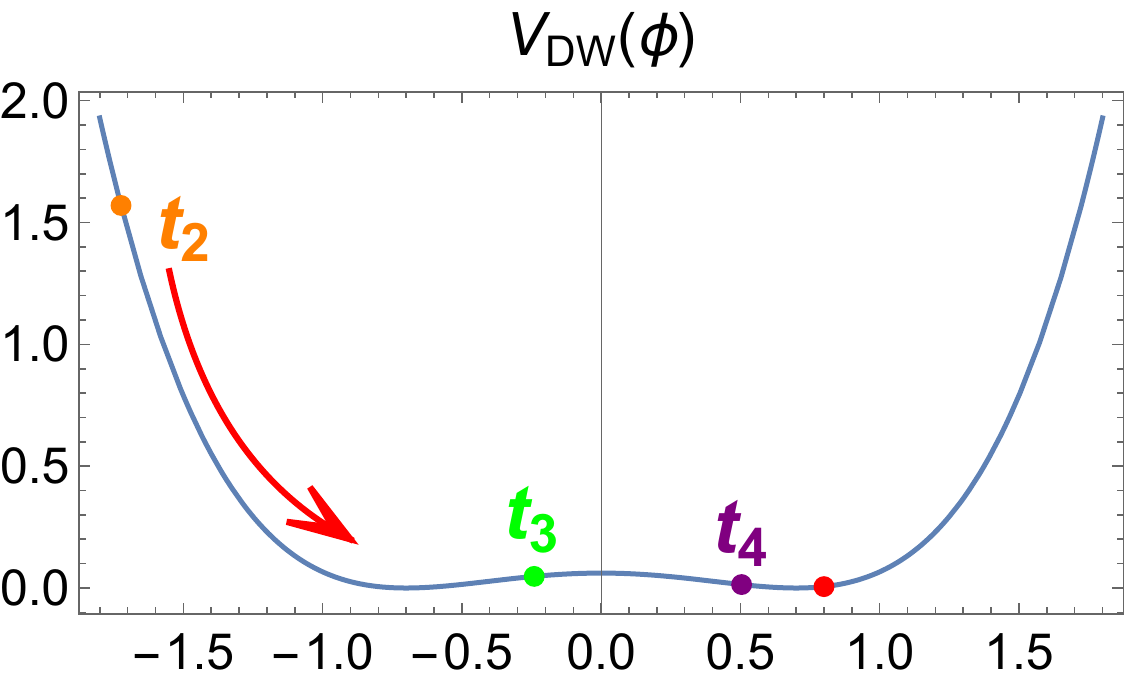}
\caption{{\bf (Left Panel)} Numerical solution for $\phi(t)\times t$ given by Eq.~(\ref{eqphi}), with $N = 60$ efolds, using the potential defined in Eq.~(\ref{VE}), with $m_\phi = 1$, $\Lambda=0$ and $a=0.7$. {\bf (Center and Right Panels)} $V[\phi(t)]$ is shown in different ranges --- notice the change of scale in both axes --- and the corresponding values of $\phi$ at the times $t_{1\to 4}$ mentioned in Fig.~\ref{RDW}. As before, the black point $t_{\rm end}$ marks the end of inflation and the red point marks the center of the last oscillatory phase in the second well (namely, $\phi=a>0$).}
\label{potphi}
\end{figure}

\begin{figure}
\center
\includegraphics[width=0.45\textwidth]{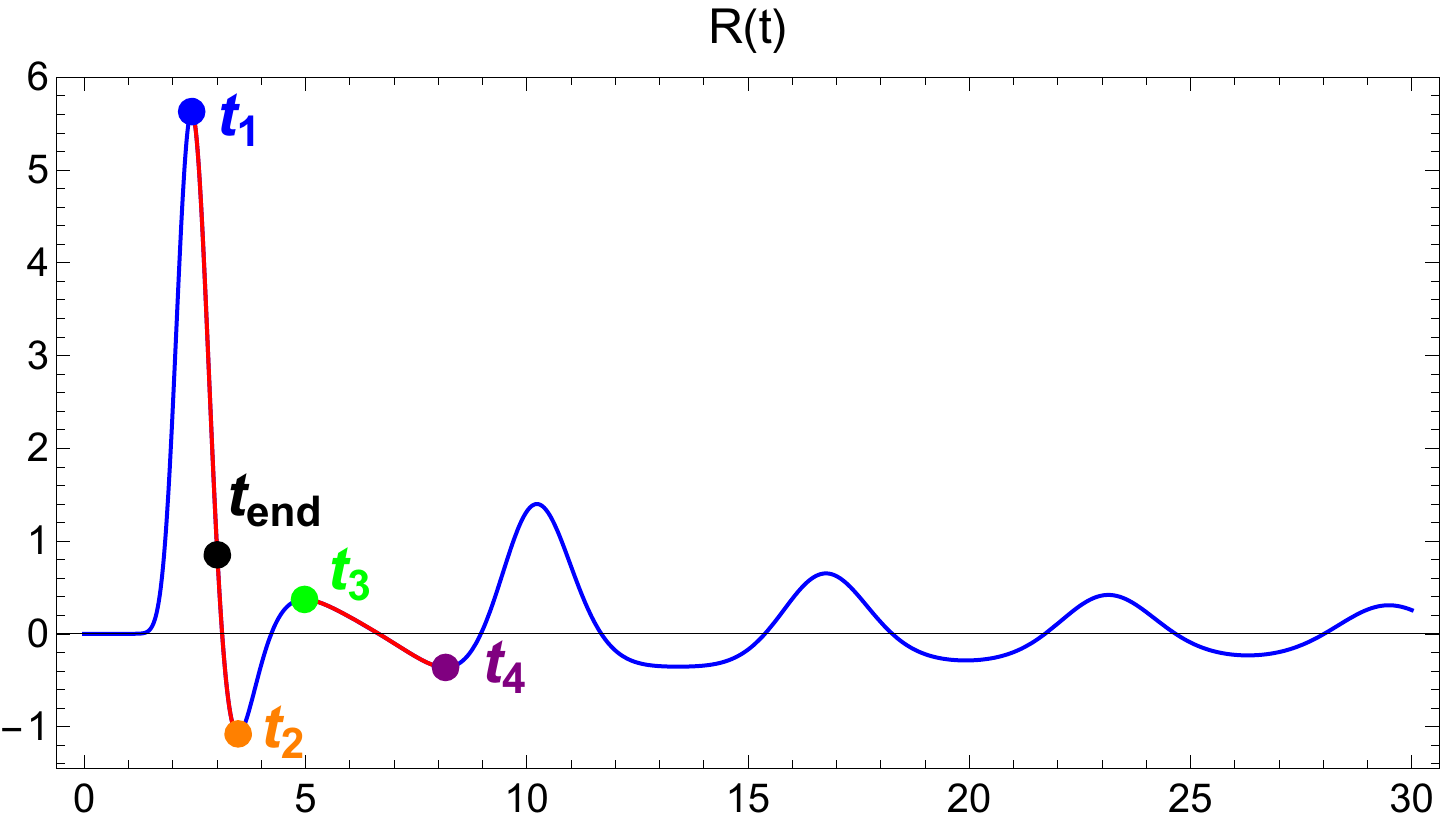}
\includegraphics[width=0.45\textwidth]{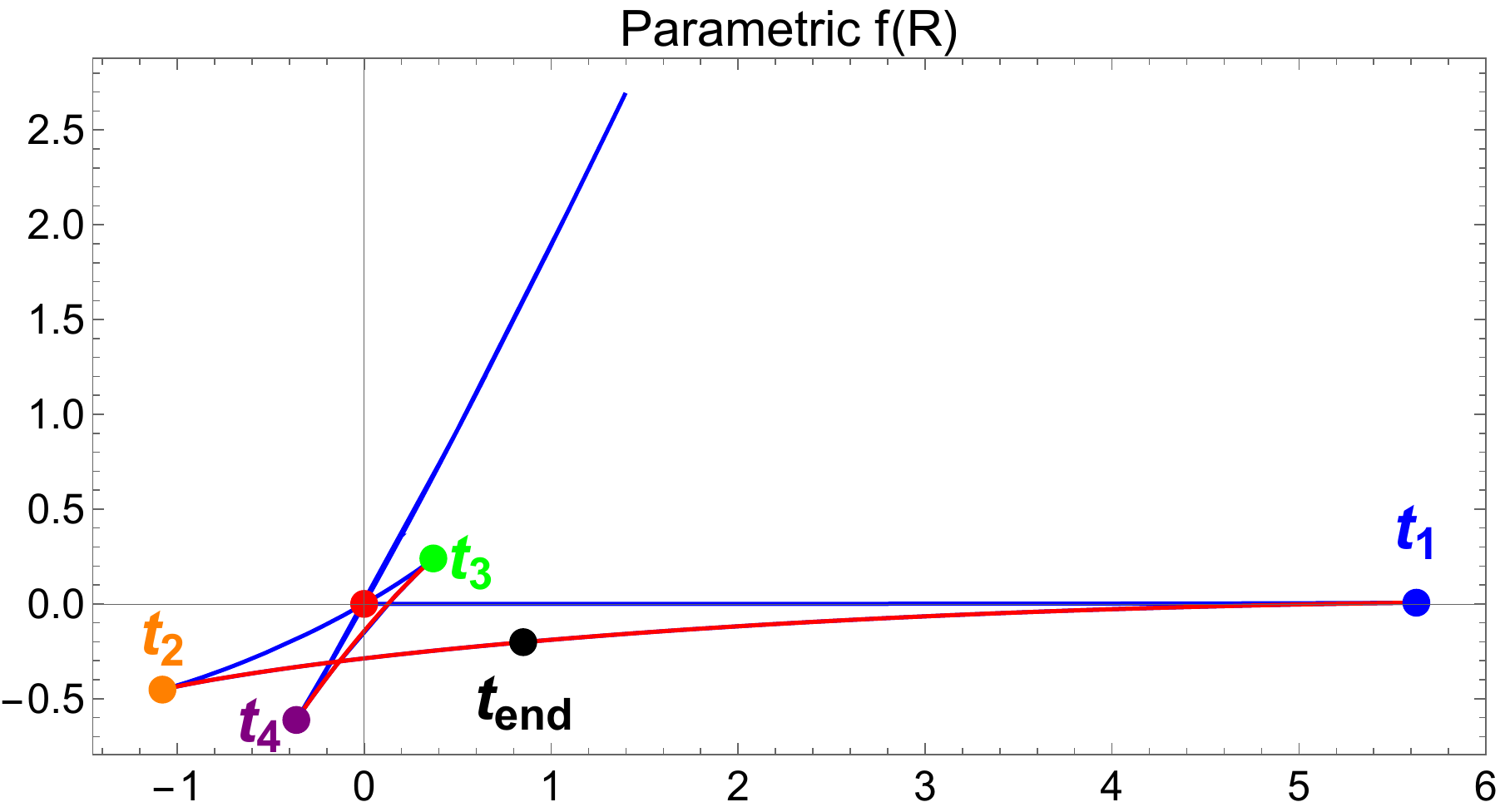}
\caption{{\bf (Left Panel)} Numerical solution for $R(t) \times t$ given by  Eq.~(\ref{RVE}), for $\Lambda=0$ and $a=0.7$, the first four extremes (two maxima and two minima) are marked with the blue, orange, green and purple points at times $t_1 = 2.42$, $t_2 = 3.47$, $t_3 = 4.97$ and $t_4 = 8.16$, respectively.  {\bf (Right Panel)} Parametric plot of the function $f[R(t)] \times R(t)$. Notice the first branch, which spreads between the origin $\{0,0\}$ and the blue point $t_1$, slightly above the horizontal axis. The first four extremes of the function $R(t)$ correspond to the four peaks of the five-branch system. The red point marks the center of the last oscillatory phase along the fifth branch. In both panels, the moment $t_{\rm end}$ indicates the end of inflation.}
\label{RDW}
\end{figure}

In Fig.~\ref{potphi} we plot the time evolution of the $\phi(t)$ field from Eq.~(\ref{eqphi}). From that piece of information and from Eq.~(\ref{RVE}), we are able to plot the numerical evolution of the $R(t)$ field in Fig.~\ref{RDW} (left panel); note the correspondence between its extrema and the cusps in the parametric plot $f(R)$ (right panel). For times $t>t_4$,  $\phi$ oscillates around $\phi=a$, which corresponds to $R=4\Lambda\exp(\beta a)$ --- we have chosen $\Lambda=0$ in Fig.~\ref{RDW}. 

\begin{figure}[t]
\center
\includegraphics[width=0.5\textwidth]{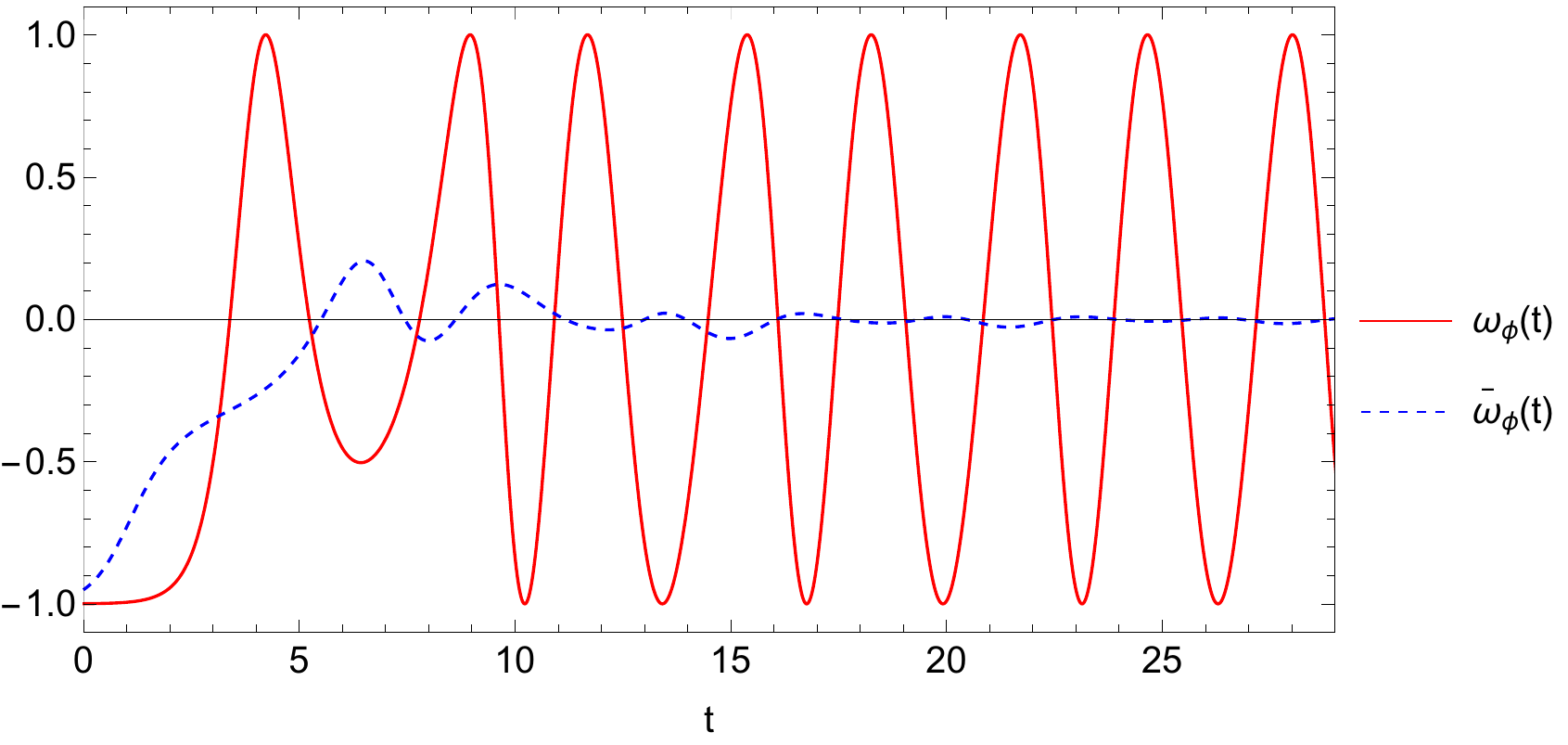}
\caption{Equation-of-state parameter ($w_\phi$ and its time average $\bar{w}_\phi$) for the $\phi$ field, defined in the EF, as functions of time, for $\Lambda=0$ $a=0.7$. Note that $\bar{w}_\phi$ can only be correctly interpreted in the oscillatory phase, where the period ${\cal T}$ can be defined.}
\label{wphi}
 \end{figure}

We plot in Fig.~\ref{wphi}, along each of the aforementioned stages, the corresponding equation-of-state parameter for the $\phi$ field (defined in the EF):
\begin{equation}
    w_\phi(t) \equiv \frac{p_\phi(t)}{\rho_\phi(t)} \equiv 
    \frac{\frac{1}{2}{\dot\phi}^2-V_E[\phi(t)]}{\frac{1}{2}{\dot\phi}^2+V_E[\phi(t)]},
\end{equation}
and its average over one period $\mathcal{T}$ (defined in the final oscillatory phase, where its calculation makes sense).
There are clearly two distinct phases: the early inflationary period, characterized by $w_\phi  \approx -1$, and the dust-like phase, when $w_\phi$ oscillates between $\pm 1$ and $\bar{w}_\phi=0$, as for the traditional inflaton field in the JF \footnote{At some point, the inflaton field should couple to matter (which is absent in our model from the beginning) to start (p)reheating --- the study of such phase is beyond the scope of the present paper.}.
\subsection{Jordan Frame (JF)}
One can describe the evolution of the system along the branches in Figs.~\ref{Butterfly1} and ~\ref{Butterfly2} for the case with $\Lambda = 0$ and $a= 0.7$, as follows: The system starts close to the origin and slowly moves along the first branch (close to the horizontal axis), generating an initial inflationary phase (since $w_\phi<-1/3$). The best fit to $f(R)$ along this first branch is $f(R) \approx R^{4.22}$, with {\bf no} GR-like term ($\propto R$). The system then quickly sweeps through the second branch (where $f''<0$) and reaches the third branch (where $f''>0$ once more) . The system continues to a fourth branch (where again $f''<0$) and then oscillates around the origin along the almost-linear fifth branch. 

On the other hand, from the extra terms in the Einstein equations, one can define a conserved ``curvature fluid'' whose energy density and pressure are, respectively:
\begin{align}
8\pi G \rho _{c} &\equiv \left( f'R - f\right)/2 - 3H\dot{f'} + 3H^{2}(1-f') \label{rhocurv}\\
8\pi G p_{c}  &\equiv \ddot{f'} + 2H\dot{f'} - (2\dot{H}+3H^{2})(1-f') + (f-f' R)/2 \label{pcurv}.
\end{align}

\begin{figure}[t]
\center
\includegraphics[width=0.4\textwidth]{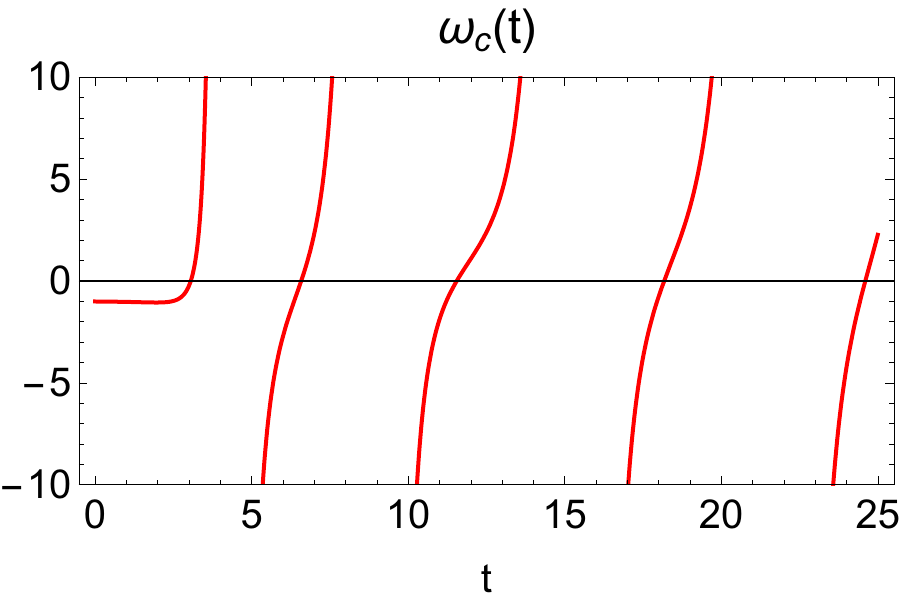}
\includegraphics[width=0.43\textwidth]{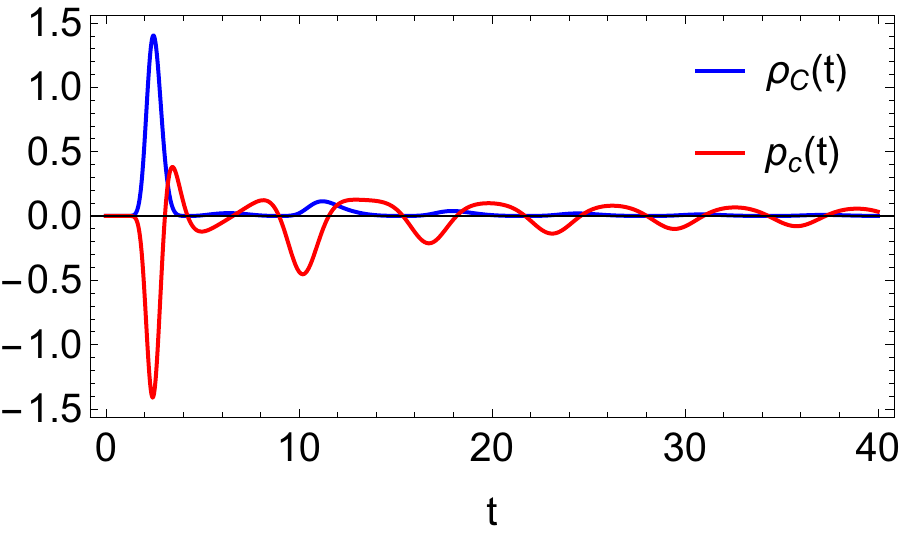}
\caption{{\bf Left panel:} Equation-of-state parameter $\omega_c$ for the ``curvature fluid'' in the JF as a function of time. The divergences, all of them non-physical, correspond to $\rho_c=0$, which happens periodically while the field $\phi$ oscillates around the minimum of its potential $V_E(\phi)$. {\bf Right panel:} Corresponding pressure $p_c$ (red solid curve) and density $\rho_c$ (blue dashed line) for the  ``curvature fluid'', as a function of time. In {\bf both panels}, $\Lambda=0$ and $a=0.7$. 
}
\label{wJF}
\end{figure}

In Fig.~\ref{wJF} we plot the corresponding equation-of-state parameter  $\omega_c \equiv p_c/\rho_c$ (left panel), and $\rho_c(t)$, $p_c(t)$ (right panel), all of them defined in the JF, for $\Lambda=0$ and $a=0.7$. In the inflationary phase, the  curvature fluid behaves as a cosmological constant ($\omega_{c}\approx -1$), as expected, since it is responsible for the accelerated quasi-de Sitter expansion. In the oscillatory phase, on the other hand, the behavior of $\omega_{c}$ diverges just because $\rho_{c}$ vanishes periodically (when $\phi(t)=a$, at the bottom of its potential $V_E(\phi)$ --- see Fig.~\ref{wJF}, right-hand panel). Nevertheless, there are no divergences of {\it physical} quantities. If $\Lambda\neq 0$, then $\omega_c = \omega_\phi = -1$ also in the final stages, as expected.

\section{Thermodynamics of \texorpdfstring{$f(R)$}{fR}}
For now, let us associate the Cosmological Constant $\Lambda$ to an effective temperature $T\equiv\Lambda$.  It is well known \cite{Figari,Gibbons}  that, in a de Sitter-like spacetime, a cosmological constant corresponds to an effective temperature due to the presence of the horizon, just like for a black hole. Thus, such correspondence comes with no surprise. We also associate the free Gibbs energy $G$ and pressure $P$ to
 \begin{align}
P &= f(R) 
\label{Pphi}\\
G &= R - 2 \Lambda e^{\beta a}.
\label{Gphi}
\end{align}
The former identification between the Lagrangian and the pressure of a given fluid is actually usual in GR. The extra term in the latter equation prevents the entropy from becoming negative, as we will see later on.

\begin{figure}[t]
\center
\includegraphics[width=0.42\textwidth]{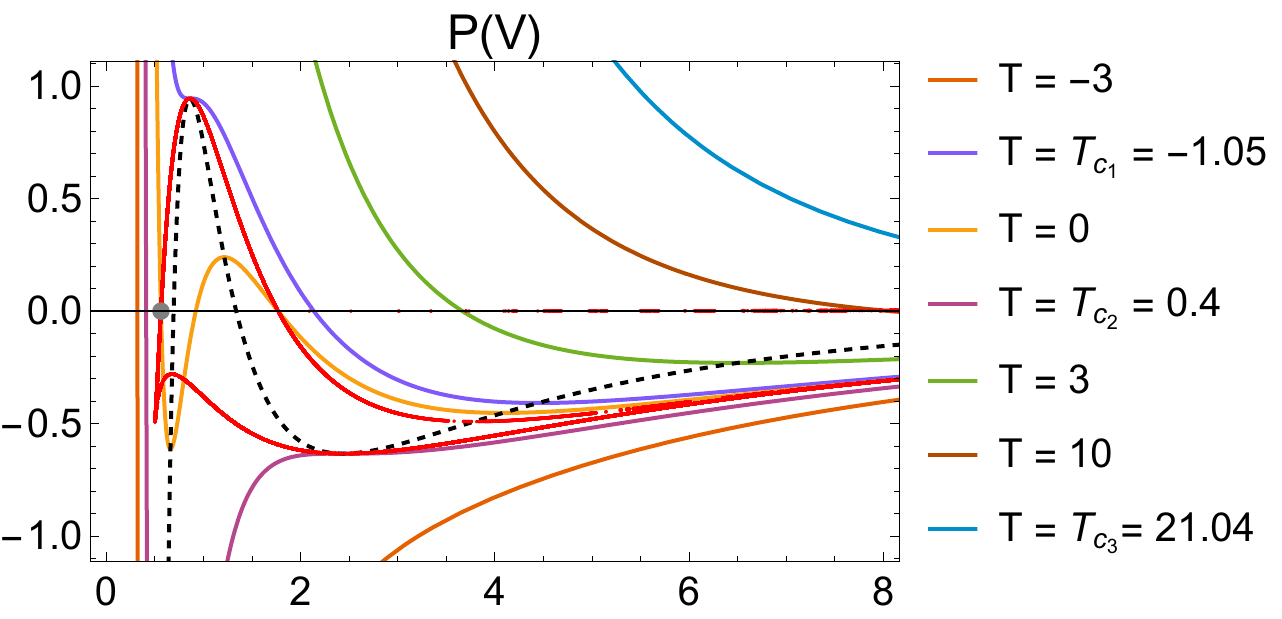}
\includegraphics[width=0.45\textwidth]{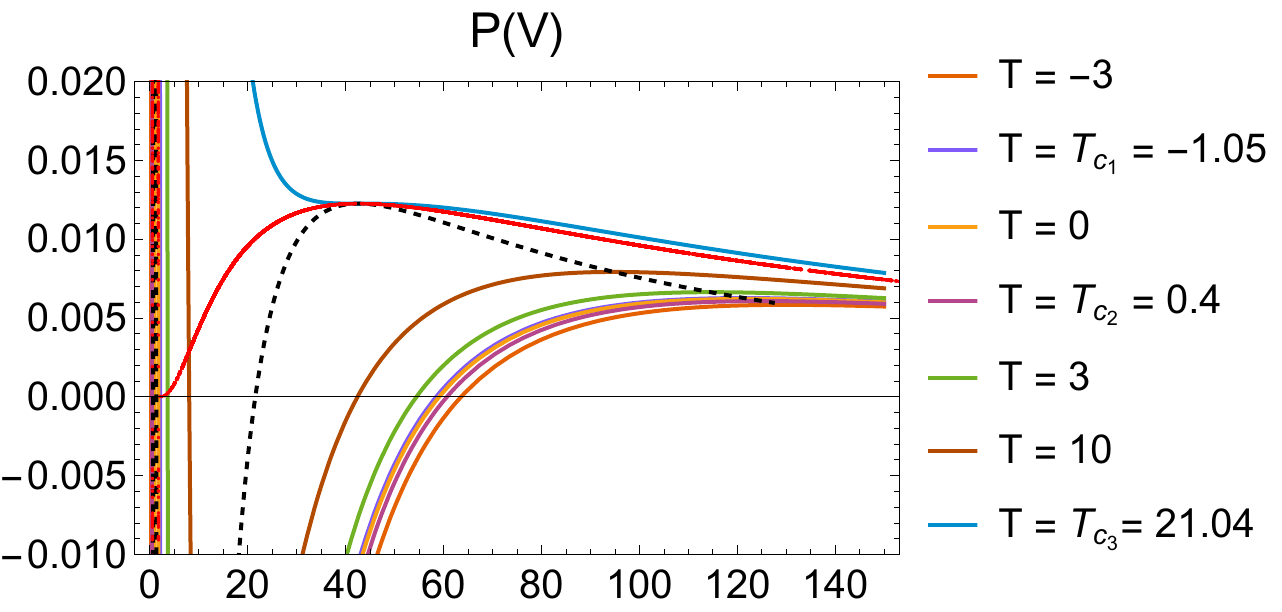}
\caption{$P \times V$ of the effective model in the $\{P,V\}$-plane, with $a=0.7$ and seven temperatures, including the three critical temperatures. The dashed black line represents the spinodal curve, the dotted red lines are the binodal curves. It is worth noting that the two panels have different scales.}
\label{PV}
\end{figure}

The effective volume $V$ is the variable ``canonically conjugated" to the effective pressure $P$, i.e, since
\begin{equation}
dG(P,T) = V \cdot  dP - S \cdot dT,
\label{dG}
\end{equation}
one can define an effective volume as
\begin{equation}
V \equiv \left.\frac{\partial G}{\partial P}\right|_T = e^{-\beta \phi} ,
\label{Vphi}
\end{equation}
which can be inverted, yielding
\begin{equation}
\phi =-\frac{1}{\beta}\ln V.
\end{equation}
Equations~(\ref{Pphi}) and (\ref{Vphi}) allow us to write the equation of state for our non-linear gas, i.e, an expression that relates $P$, $V$ and $T$:
\begin{equation}
P = \frac{1}{4 a^2 \beta^4 V^2}\left[m_\phi^2 \left(\ln^2 V-a^2 \beta ^2\right)^2-\left(\ln^2 V-a^2 \beta ^2\right)4 m_\phi^2 \ln V+8 a^2 \beta ^4 T\right].
\label{PVeq}
\end{equation}
The behaviour of $P(V)$ is shown in Fig.~\ref{PV} for a couple of values of $T$, which bears some resemblance to a vdW gas, which does have a stronger similarity to the single-well problem we have studied before \cite{Peralta}\footnote{Here one obtains $P\propto T V^{-2}$ in the high-temperature limit, instead of the standard ideal-gas behavior $P \propto T V^{-1}$.}. 
In spite of such similar curves $P(V$), as we will see in a moment, there is a plethora of new phenomena in our gas, such as three critical temperatures.

\begin{figure}[t]
\begin{minipage}{6in}
\center
\includegraphics[align=c,width=0.45\textwidth]{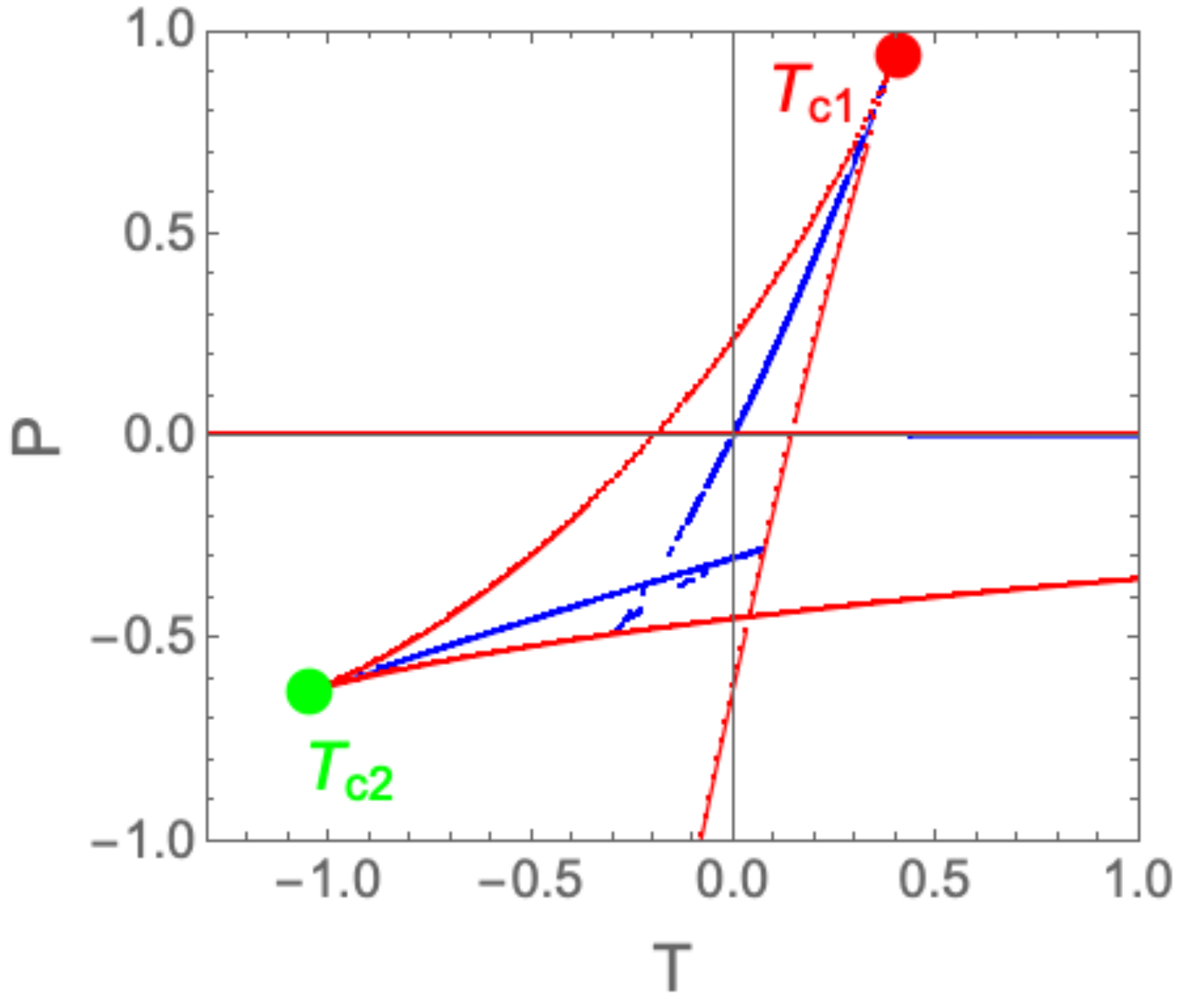} 
\includegraphics[align=c,width=0.45\textwidth]{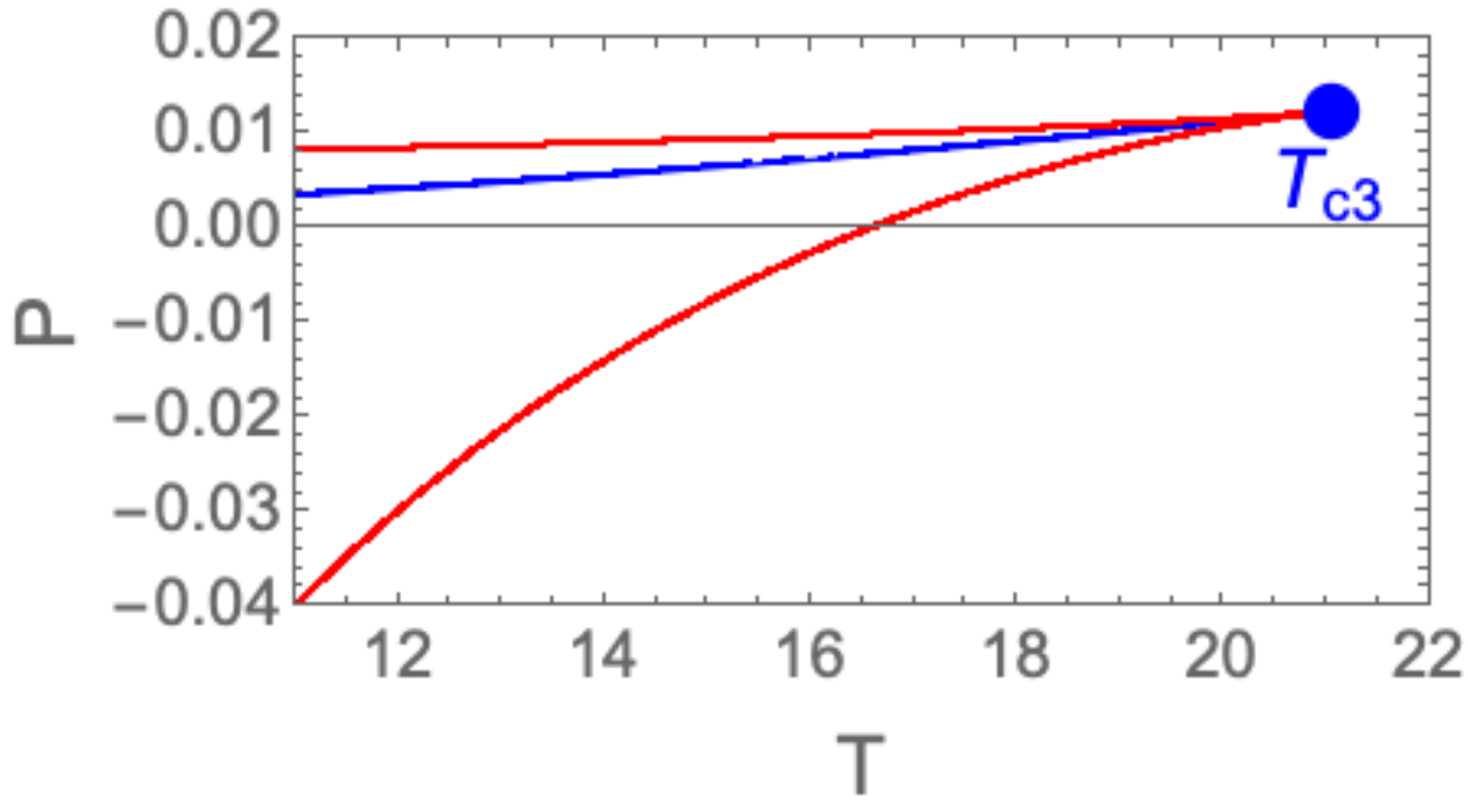}
\end{minipage}
\caption{Phase diagram for the effective model in the $\{P,T\}$-plane for $a=0.7$. The binodal curve (solid blue with gaps due to numerical errors), ending at the three critical points $T_{c_1}$, $T_{c_2}$, $T_{c_3}$ in red, green, and blue dots respectively. The spinodal curves are shown as solid red lines. The right-hand panel is a zoom in the rightmost part of the phase diagram (not shown in the former panel); note the different axis range.} 
\label{sbin}
\end{figure}

The magnitude of  the qualitative  differences from a standard vdW gas can be easily seen when we plot the spinodal and the binodal (or coexistence)  curves, which indicate, respectively, the regions of instability and metastability of the system --- see Fig.~\ref{sbin}. The former curve (spinodal) is obtained either from  the lateral extrema (``wings") of the Gibbs function (see Fig.~\ref{RDW}), i.e, the first four turning points (the global and first three local extrema) of $R(t)$ or from the extrema of the $P\times V$ plot, where $dP/dV=0$. The latter curve (binodal) can also be obtained using two equivalent calculations: from the self-intersecting points of the Gibbs function (plotted as a function of the pressure, for fixed temperature) and from the Maxwell construction, supporting the results from each other. The three {\it critical points} $\{P_c,T_c,V_c\}$ are defined at the crossing of each pair of those curves. For the case of $\Lambda = 0$ and $a =0.7$, the system features three critical temperatures, namely $T_{c1}$, $T_{c2}$ and $T_{c3}$ shown in Fig.~\ref{sbin}. See also Fig.~\ref{Butterfly2} to follow the evolution of the Gibbs function as the temperature ($\Lambda$) increases. 

If we take $T = 0$ (solid orange curve in Fig.~\ref{PV}), the system starts at $V\to\infty$ in a metastable phase (the binodal region) --- the initial inflationary solution is indeed momentary.
Either way, the effective fluid quickly crosses the spinodal curve (the unstable region) and then oscillates around  $P=2 T \exp(2 \beta a)$ and $V=\exp(-\beta a)$, indicated by a gray circle. At this temperature, the system ends exactly on the binodal curve. For higher temperatures, though, the system settles down above the binodal line, i.e, in a stable configuration.

One can also calculate the Helmholtz energy
\begin{align}
F(T,\phi) &\equiv G - P \cdot V = \frac{e^{\beta  \phi }}{4 a^2} \left(a^4 m^2_{\phi}+a^2 \left(8 T-2 m^2_{\phi} \phi ^2\right)+m^2_{\phi} \phi ^4\right)-2 T e^{a \beta },
\end{align}
from which one can define the entropy as
\begin{align}
S (T,\phi) &\equiv - \left.\frac{\partial F}{\partial T}\right|_\phi = \frac{1}{2} \left(4 e^{a \beta }-4 e^{\beta  \phi }\right).
\end{align}
One can then realize that the specific heat at constant volume vanishes, since $C_V \equiv T \cdot \partial S/\partial T|_V  = 0 \, \forall T$. Such feature is not unusual: it has been already found in studies of  thermodynamics and phase transitions of black holes \cite{Dolan_2011}.

Another important feature is the sudden change in entropy, from $S(\phi\to-\infty)=2 \exp{(\beta a)}$  to $S(\phi=a)=0$, marking the release of latent heat, just as expected in an ordinary first-order phase transition (which will be shown below by the $C_P$ behaviour). Such spontaneous decrease in entropy correctly indicates that the gravitational sector described in this paper is incomplete. we recall the reader that we have completely neglected the matter sector from the beginning, i.e, there is no energy-momentum tensor for the matter sector. The transfer of energy between those sectors is the well-known (p)reheating mechanism, which will be the subject of future work.

\begin{figure}[t]
\center
\includegraphics[width=0.4\textwidth]{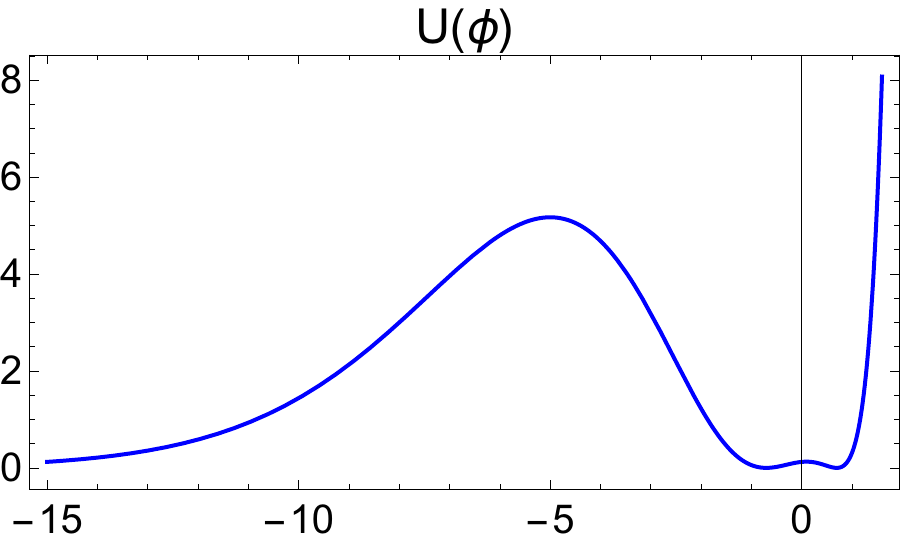}
\includegraphics[width=0.4\textwidth]{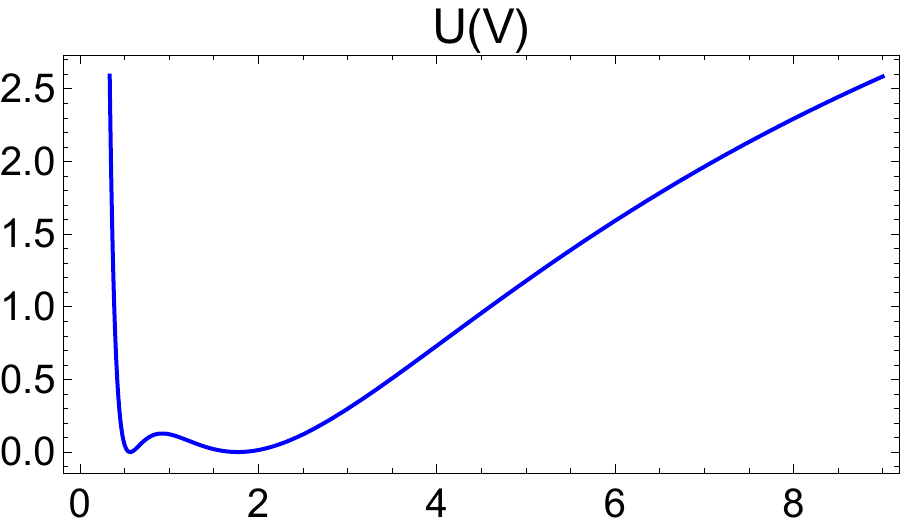}
\caption{Plot of the \textcolor{black}{internal energy} $U$ as a function of $\phi$ (left panel) and of the volume $V$ with $a=0.7$ (right panel). The system starts on a stable (asymptotic) solution $\phi\to-\infty$ ($V\to+\infty$), but the slow roll  drives the field towards the origin. Eventually, it settles down at the minimum $\phi=a$ ($V=\exp(-\beta a)$).}
\label{Ufigs}
\end{figure}

The internal energy $U(T,\phi)$ is given by its standard definition:
\begin{align}
U &\equiv G - P \cdot V + T \cdot S = \frac{m_\phi^2 }{4 a^2}\left(a^2-\phi ^2\right)^2 e^{\beta  \phi }
\end{align}
for which $\phi=a$ (or, accordingly, $V=\exp(-\beta a)$) is always a minimum. It turns out that also $U$ is only a function of the volume $V$ and {\it not} of the temperature $T$. We acknowledge  (see Fig.~\ref{Ufigs}) the existence of two ``extra" equilibrium points --- a stable asymptotic one (a minimum at $\phi\to-\infty$) and a local maximum (at $\phi = (-2\pm\sqrt{a^2 \beta ^2+4})/\beta$) --- besides the expected ones at $\phi=\pm a$. 

\begin{figure}[t]
\center
\includegraphics[width=0.3\textwidth]{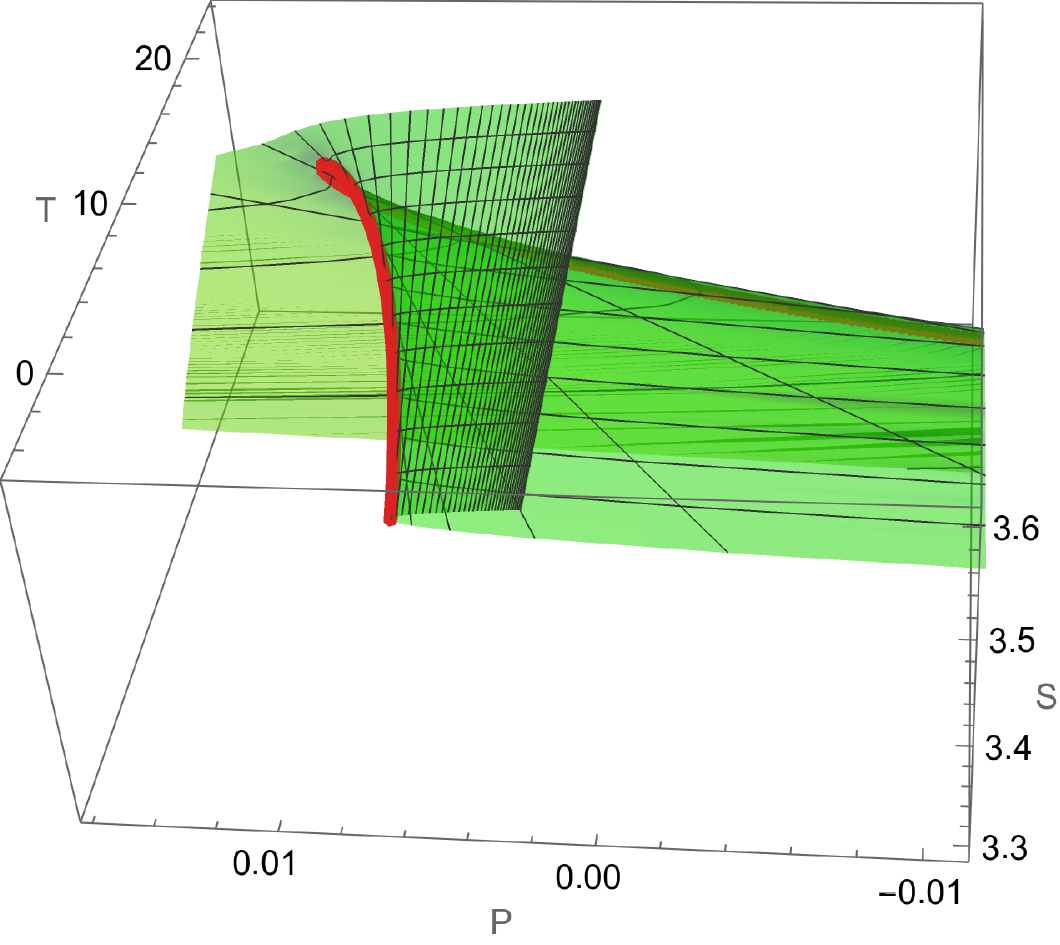}
\includegraphics[width=0.3\textwidth]{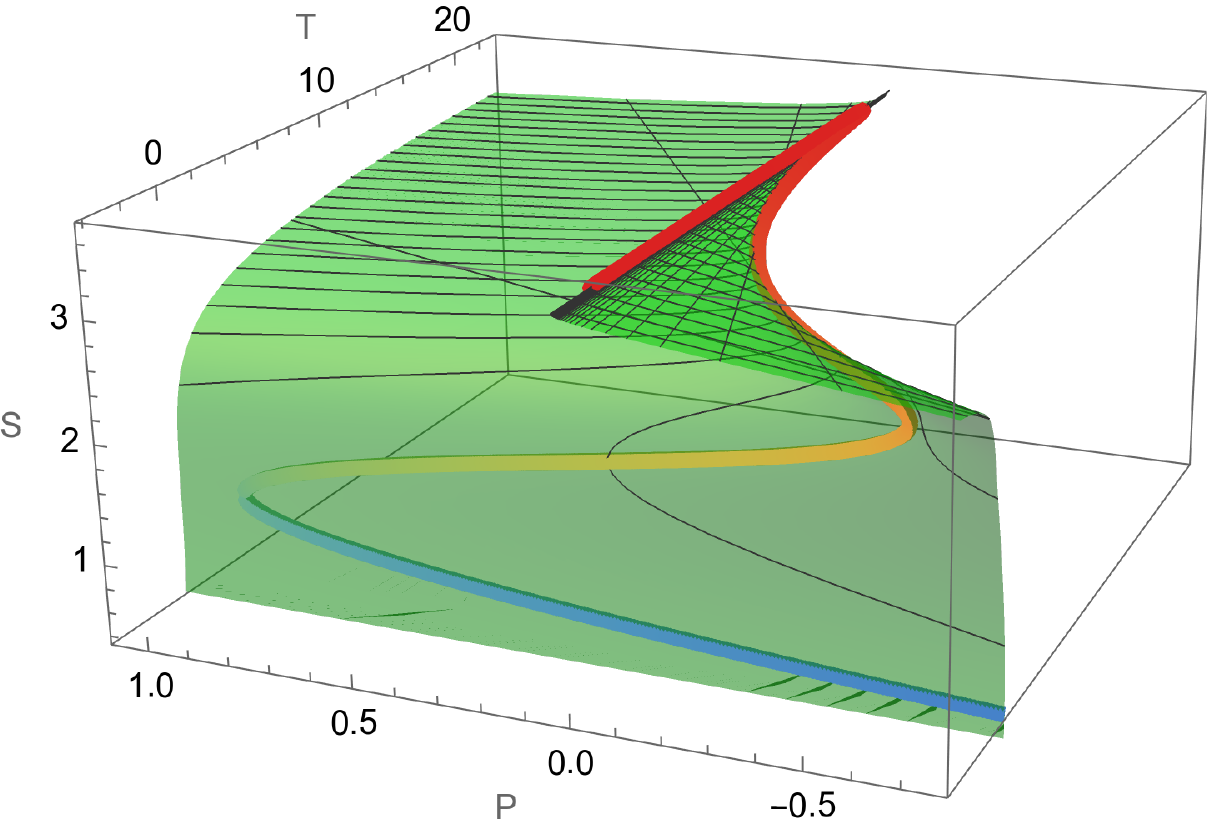}
\includegraphics[width=0.3\textwidth]{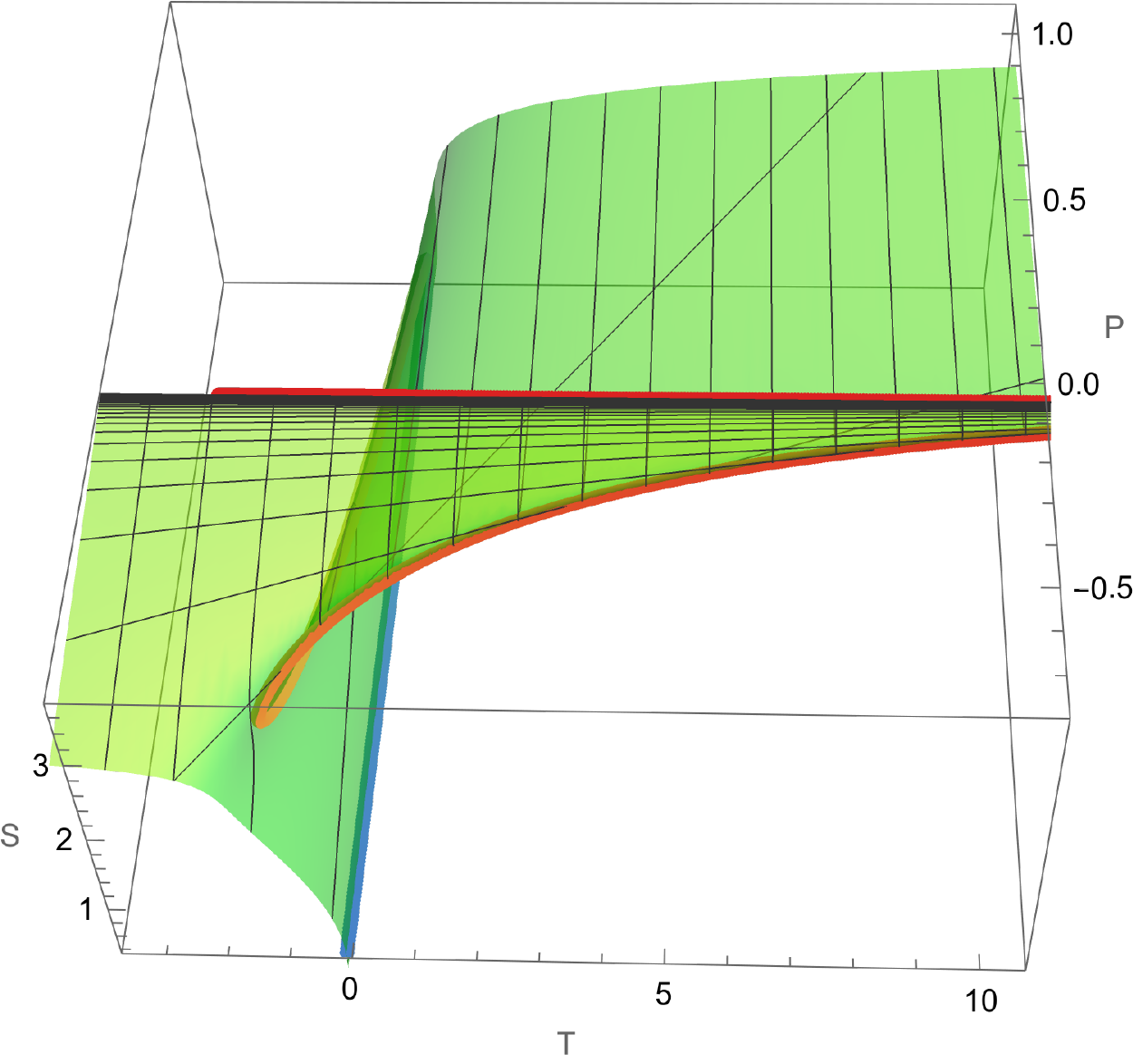}
\caption{Entropy surface given by $S(P,T)$ for $a=0.7$ and $\Lambda = 0$ (All panels) for differents pressures scales. For this case the surface presents a triple fold. The spinodal curve is indicated in color rainbow function..}
\label{spt}
\end{figure}

The entropy as a function of pressure and temperature provides another very important piece of information.  $S(P,T)$ is depicted in Fig.~\ref{spt}, which also shows the spinodal and binodal curves.  The region where the entropy is multi-valued is known in Catastrophe Theory \cite{saunders1980introduction} as a cusp and indicates the existence of a first-order phase transition and unstable configurations.

\begin{figure}[t]
\center
\includegraphics[width=0.3\textwidth]{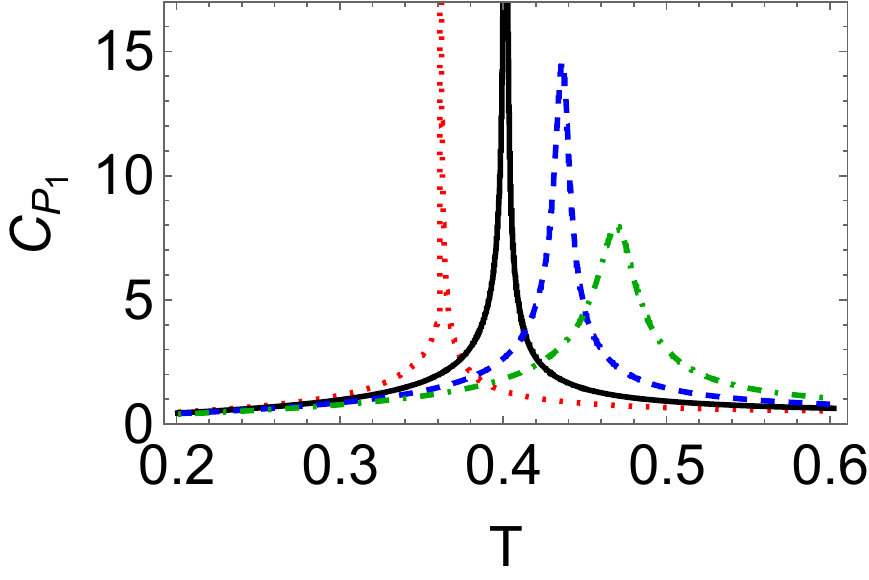}
\includegraphics[width=0.3\textwidth]{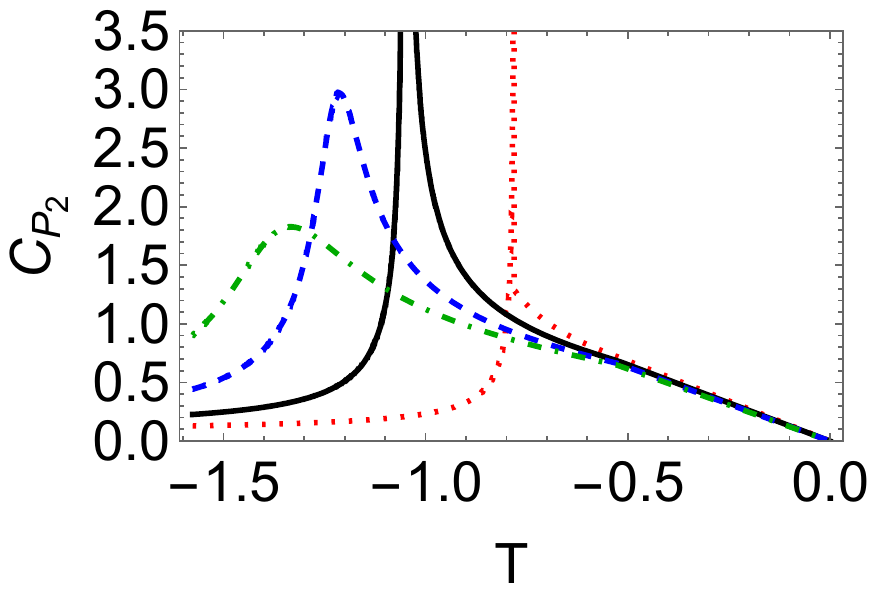}
\includegraphics[width=0.3\textwidth]{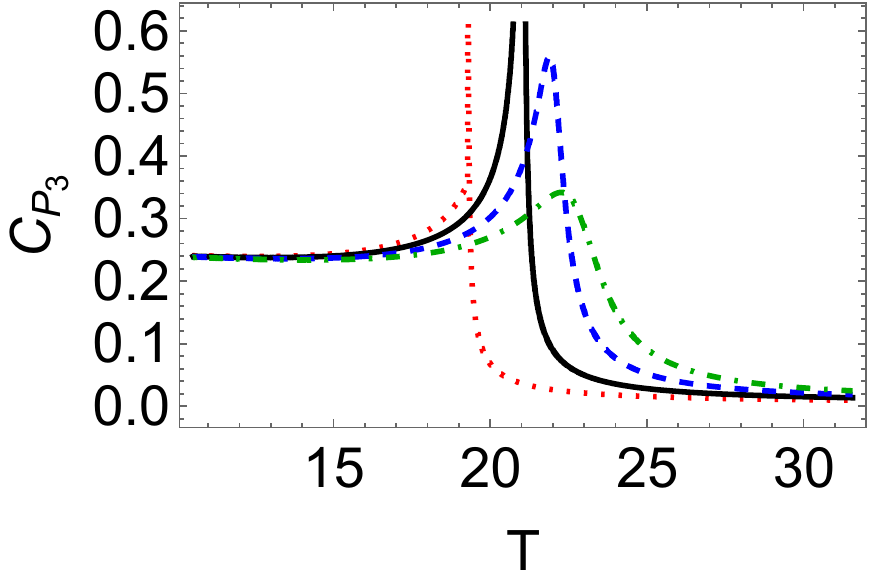}
\caption{Behavior of the specific heat at constant pressure $C_P$ as a function of the temperature $T$ close to its transition values: $T_{c1}=0.4016$ if $P=P_{c1}=0.9447$ (left panel), $T_{c2}=-1.0507$ if $P=P_{c2}=-0.6332$ (center panel), $T_{c3}=21.0430$ (right panel) if $P=P_{c3}=1.2252\times 10^{-2}$, for different values of pressure: $0.9 P_c$ (dotted red), $P_c$ (solid black), $1.1 P_c$ (dashed blue) and $1.2 P_c$ (dot-dashed  green). In all curves, $a=0.7$.}
\label{Cp}
\end{figure}

From $S(P,T)$ we can get the specific heat at constant pressure, $ C_P \equiv T \cdot \partial S/\partial T |_P$, shown in Fig.~\ref{Cp}. We obtain the expected behavior for temperatures around the coexistence curve, for pressures both below (finite jump) and above (smooth behavior) the critical value $P_c$. We also obtain the usual divergence at the critical point $\{T_c,P_c\}$ (solid black line in Fig.~\ref{Cp}) as given by $C_{P}|_{P_{c_1}} \sim [(T-T_{c_1})/T_{c_1}]^\alpha$, with $\alpha\approx 1.36$, $C_{P}|_{P_{c_2}} \sim [T-(T_{c_2})/T_{c_2}]^\alpha$, with $\alpha\approx 1.39$, $C_{P}|_{P_{c_3}} \sim [(T-T_{c_3})/T_{c_3}]^\alpha$, with $\alpha\approx 18.92$. 
\begin{figure}[!t]
\center
\includegraphics[width=0.5\textwidth]{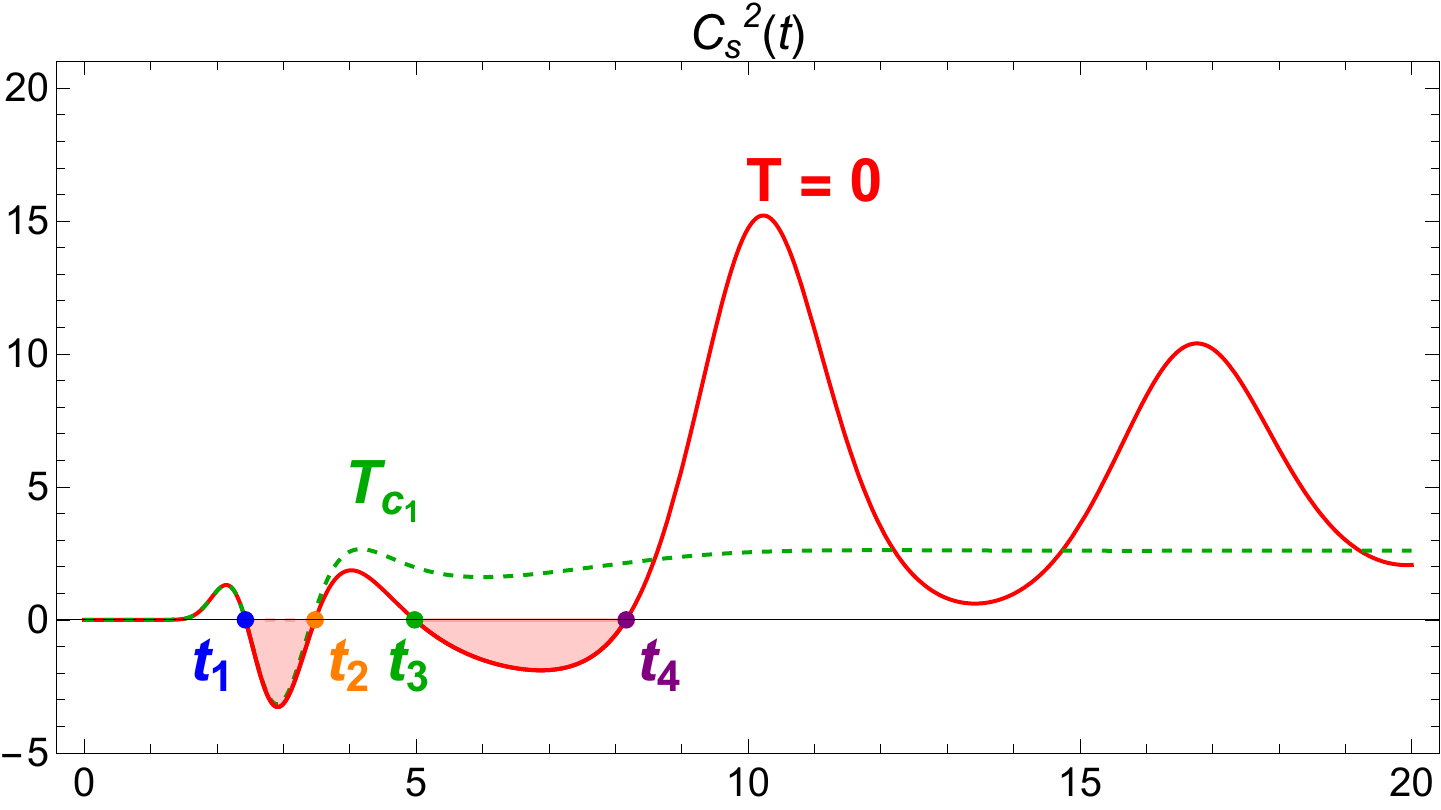}
\caption{Plot of  $\kappa \cdot c_{\rm s}^2 \equiv \kappa \cdot \dot P/ \dot \rho$ for the non-linear gas for $a=0.7$. $T=0$ (red), and $T=T_{c_1}$ (dotted green), as functions of time. The dots indicate when $\dot R(t)=0$, i.e, at the sideway peaks in Fig.~\ref{RDW}, between which $f''(R)<0$.}
\label{cs2}
\end{figure}
The sound speed squared, defined as $c_{\rm s}^2 \equiv \dot P/\dot \rho = -(V^2/\kappa) \dot P / \dot V$ (where we define $\kappa>0$ by $\rho =: \kappa /V$) and  plotted in Fig.~\ref{cs2}. We can see that $c_{\rm s}^2<0$ {\it only} between the first two (and  between the third and fourth) extrema of $R(t)$, i.e, in the second and the fourth branches (see Fig.~\ref{RDW}), when $f''<0$, as expected from the usual {\it perturbative} argument on stability of $f(R)$ theories \cite{Sotiriou:2008rp}.  
With an imaginary sound speed, fluctuations grow exponentially fast, but, during the spinodal decomposition process, only a given range of wavelength do so \cite{CHOMAZ2004263}. This is similar to a feature that has already been proposed in the preheating scenario \cite{preheating}. Further details will be the subject of future work. 

\section{Conclusions}


In this paper, we have investigated the $f(R)$ theories of gravity where $f(R)$ is a nonlinear functions of the Ricci scalar $R$ in the Jordan Frame, using the metric formalism which features an extra degree of freedom. We have focused on the inverse problem, mapping the Einstein Frame Lagrangian onto the corresponding Lagrangian in the Jordan Frame for a double well potential with an ad-hoc Cosmological Constant $\Lambda$. We have found that the evolution of the system in the later frame occur along various branches of the $f(R)$ function according the configuration of the initial conditions of the scalar field and the dynamics of the free parameters $\Lambda$ and $a$. 

\begin{figure}
\center
\includegraphics[width=0.4\textwidth]{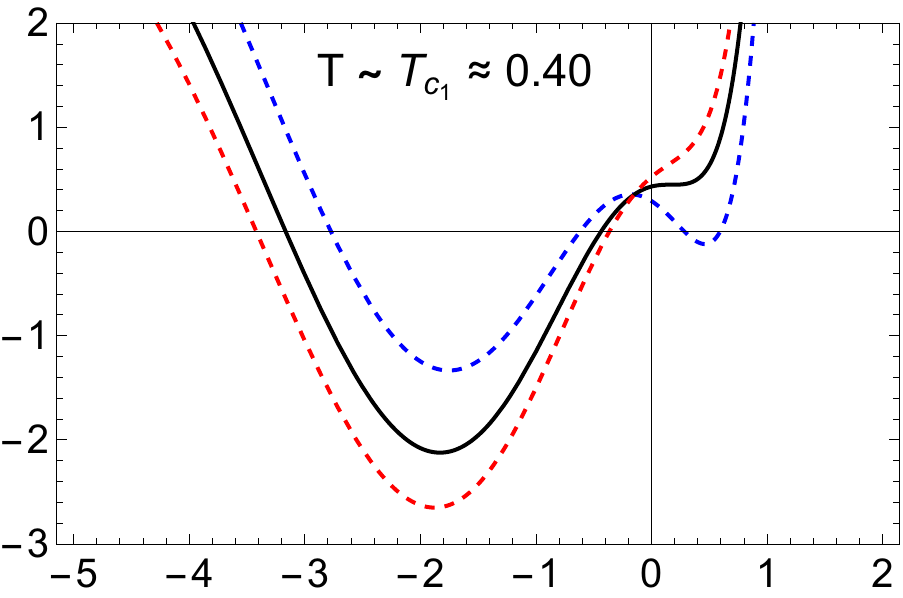}
\includegraphics[width=0.4\textwidth]{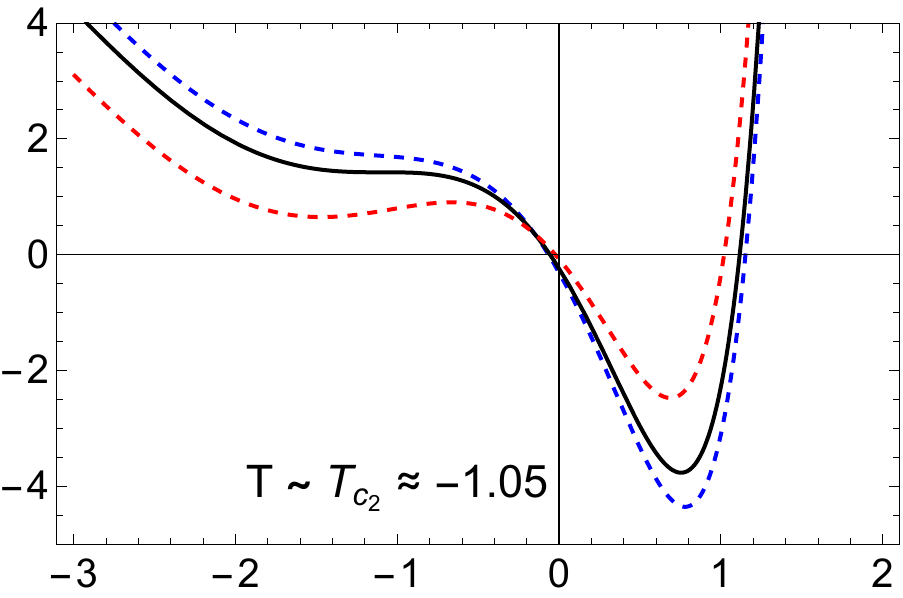}
\includegraphics[width=0.4\textwidth]{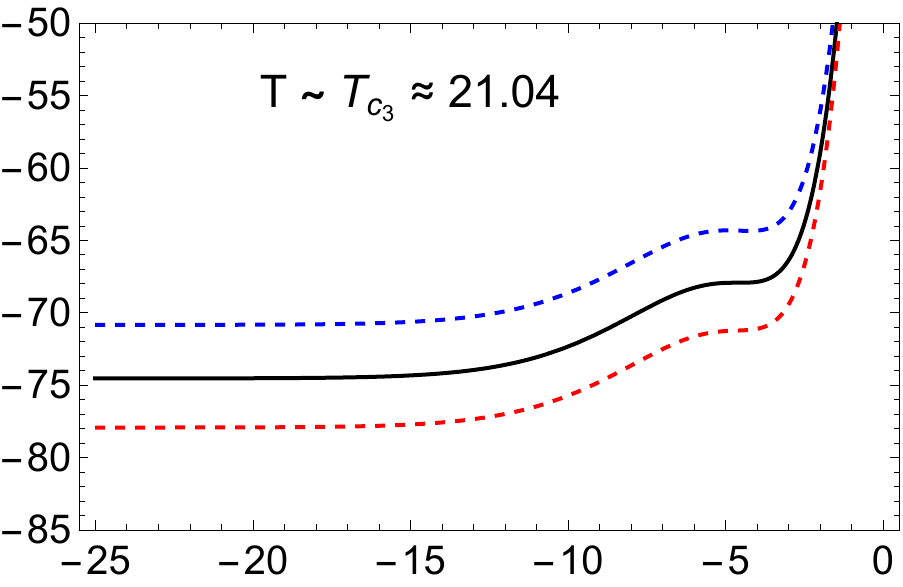}
\caption{Plots of $G \times \phi$ (where $a=0.7$) for temperatures around the critical temperatures $T_{c_1}\approx 0.4$ ({\bf top left}), $T_{c_2}\approx -1.05$  ({\bf top right}) and $T_{c_3} \approx 21.04$  ({\bf bottom}). The exact variation was chosen so that the extrema (or lack thereof) could be easily seen. In all plots, the black solid curve corresponds to the exact critical temperature, the blue-dashed line corresponds to a slightly lower temperature, and the red-dotted line to a slightly higher one. In the last plot, the curves were vertically displaced so that they all fall into the same $G$ range. Notice also that in the second case, where the critical temperature is negative, the {\bf higher}-temperature curve features the two new extrema (besides the persistent one at $\phi \sim 1$, which is annihilated only at $T_{c1}$), which the opposite behaviour of a vdW-like fluid.}
\label{extrema}
\end{figure}

We have explored the thermodynamics interpretation for this case, where the cosmological constant is associated with an effective temperature, the free Gibbs energy to the Ricci scalar $R$ and pressure to the Lagrangian in the jordan frame and derived the effective volume as the variable conjugate to the effective pressure. We have also derived an equation of state for our non-linear gas that relates pressure, volume, and temperature. We have shown that the behavior of the pressure-volume curve bears some resemblance to a van der Waals gas, but exhibits a variety of new phenomena, such as three critical points.
The three critical temperatures and their related pairs of spinodal and binodal lines correspond to three first-order phase transitions. Indeed, the Gibbs Potential does present the expected coalescence of extrema when plotted as a function of $V$ (or  $\phi$, which features a nicer scale range) at each $T_{ci}$, as shown in Fig.~\ref{extrema}. For each $T_{ci}$, there is one value of $P_i$; all of them were already indicated in Fig.~\ref{sbin}. For temperatures (higher) lower than ($T_{c2}$) $T_{c1}$ and $T_{c3}$, there is a line (binodal) in the phase diagram where two phases can coexist. The crossing of the binodal lines (at about $P_*\approx -0.36$, $T_*\approx -0.20$) indicates a ``triple" point, where {\bf all} phases coexist. 

In addition to calculating standard thermodynamic quantities like the Helmholtz energy, internal energy, entropy, specific heats at constant volume and constant pressure, and sound speed squared, our comprehensive approach also sheds light on the evolution of the system in the Jordan frame. By exploring the behavior of $f(R)$ theories in relation to thermodynamics, we offer a unique perspective on the thermodynamics of spacetime.

\section*{Acknowledgements}
CDP thanks Yeinzon Rodr\'iguez for his support during the developed of this research and acknowledges financial support from MinCiencias, Colombia, under the program “estancias posdoctorales convocatoria 891-2020”, grant number: 80740-687-2021, and Centro de Investigaciones en Ciencias B\'asicas y Aplicadas at Universidad Antonio Nari\~no. SEJ thanks Eduardo Fraga for insightful discussions on the subject and FAPERJ for the financial support.
   
%

\end{document}